\documentclass[12pt,reqno]{amsart}
\usepackage{graphicx}
\usepackage{amssymb,amsmath}
\usepackage{amsthm}
\usepackage{color}
\usepackage[pdf]{pstricks}
\usepackage{hyperref}
\usepackage{empheq}
\usepackage{pgfplots}
\usepackage{subcaption}
\usepackage{extarrows}
\usepackage{calc}

\newtheorem{remark}{Remark}
\newtheorem{lemma}{Lemma}

\newtheorem{prop}{Proposition}
\newtheorem{theorem}{Theorem}
\newtheorem{definition}{Definition}
\newtheorem{example}{Example}

\numberwithin{equation}{section}

\newcommand{\beq}{\begin{equation}}
\newcommand{\eeq}{\end{equation}}
\newcommand{\ben}{\begin{eqnarray}}
\newcommand{\een}{\end{eqnarray}}
\newcommand{\beno}{\begin{eqnarray*}}
\newcommand{\eeno}{\end{eqnarray*}}

\allowdisplaybreaks

\newdimen\eqjot \eqjot = 1\jot
\def\openupeq{\openup \the\eqjot}

\begin{document}
\title[Algebraic dark solitons versus rogue waves in MTM]{Algebraic dark solitons versus rogue waves \\ in the massive Thirring model}
	
\author{Zhen Zhao}
\address[Z. Zhao]{School of Mathematics and Statistics, Ningbo University, Ningbo 315211, People's Republic of China and Department of Mathematics and Statistics, McMaster University, Hamilton, Ontario, Canada, L8S 4K1}
\email{zhaozhen00728@163.com}
	
\author{Dmitry E. Pelinovsky}
\address[D. E. Pelinovsky]{Department of Mathematics and Statistics, McMaster University, Hamilton, Ontario, Canada, L8S 4K1}
\email{pelinod@mcmaster.ca}
	
\begin{abstract}
We address the existence of algebraically decaying solitary waves on the nonzero wave background in the massive Thirring model (MTM) in laboratory coordinates. We find the precise condition of the spectral stability and instability of the nonzero wave background. In the case of instability, we distinguish modulational, side-band, and full-band instabilities and show that the rogue wave exists if and only if the nonzero wave background is modulationally unstable. For the other unstable cases and in the case of stability, we show that there exist exactly two algebraic dark solitons which correspond to the end points in the admissible range of the wave speed for the dark soliton propagation. 
\end{abstract}   
	
	\maketitle
	
	
\section{Introduction}

\subsection{The model}

A relativistically invariant nonlinear Dirac equation was proposed in the quantum field theory by W. Thirring in \cite{Thirring-AOP-1958}. The local self-interaction model of a Dirac field, which is now referred to as the massive Thirring model (MTM), can be written in one spatial dimension as follows:
\begin{equation}
\label{MTM-original}
	\begin{cases}
		{\rm i} \psi_t - \phi_x - \psi = (\psi^2 + \phi^2) \bar{\psi}, \\
	{\rm i} \phi_t + \psi_x + \phi = (\psi^2 + \phi^2) \bar{\phi},
	\end{cases}	
\end{equation}
where $\psi$ and $\phi$ stand for two components of the spinor field. 
With the transformation 
\begin{equation*}
	\begin{cases}
\psi = \frac{1}{2} (u - v), \\
\phi = -\frac{{\rm i}}{2}(u + v),
 	\end{cases}	
 \end{equation*}
the system (\ref{MTM-original}) can be rewritten in the equivalent form
\begin{equation}
	\label{MTM}
	\begin{cases}
		{\rm i} (u_t+u_x)+v = |v|^2 u, \\
		{\rm i} (v_t-v_x)+u = |u|^2 v.
	\end{cases}
\end{equation}
This model appears in the numerous applications as a limiting case 
of the coupled-mode system for two counter-propagating waves coupled due to the subharmonic (Bragg) resonance, e.g., in the nonlinear wave models with a periodic potential \cite{SS94} (see book \cite{Pel11}). 

\subsection{The problem and background literature}

The MTM in the form (\ref{MTM}) was shown to be integrable with the Lax pair and infinitely many conserved quantities, see \cite{Kuznetsov-TMP-1977,Mikhailov-JETP-1976} and  \cite{Kaup-ANC-1977,Orfanidis-PRD-1976}. Multi-soliton solutions with the exponential spatial decay to the zero background were constructed in \cite{Bar1,Date,DHS}. These solutions are usually referred to as the bright solitons. Multi-soliton solutions with the exponential spatial decay to the nonzero wave background (referred to as the dark solitons) were constructed in \cite{Bar2,Bar3}. Both bright and dark soliton solutions were recently represented by using Hirota’s bilinear method and the KP-hierarchy reductions in \cite{Chen-SAPM-2023}. Generalizations of bright and dark soliton  solutions for the integrable discretizations of the MTM system were developed recently in \cite{CF1,CF2}. Additionally, quasi-periodic solutions were constructed by using the Baker--Akhiezer wave function and the Riemann theta function in \cite{Bikbaev}, see also \cite{Date2,Wisse}.

The relevance of both bright and dark solitons can be understood by reducing the MTM to the nonlinear Schr\"{o}dinger (NLS) equations in the nonrelativistic limit. On the zero background, the spectrum of the linear self-adjoint Dirac operator, 
\begin{equation}
\label{Dirac}
D : H^1(\mathbb{R}) \subset L^2(\mathbb{R}) \to L^2(\mathbb{R}), \quad 
D = \left( \begin{matrix} -{\rm i} \partial_x & -1 \\ -1 & {\rm i} \partial_x \end{matrix} \right)
\end{equation}
is purely continuous, located on $(-\infty,-1] \cup [1,\infty)$. 
Near the end points of the spectral gap $(-1,1)$, 
we proceed with the asymptotic multiple-scale expansion of 
the corresponding solutions to the nonlinear system (\ref{MTM}),
\begin{equation*}
\begin{cases}
u(x,t) = e^{\pm {\rm i} t} [\varepsilon A_{\pm}(X,T) \pm \frac{{\rm i}}{2} \varepsilon^2 \partial_X A_{\pm}(X,T) + \varepsilon^3 u_3(X,T) + \mathcal{O}(\varepsilon^4)], \\
v(x,t) = e^{\pm {\rm i} t} [ \pm \varepsilon A_{\pm}(X,T) - \frac{{\rm i}}{2} \varepsilon^2 \partial_X A_{\pm}(X,T) + \varepsilon^3 v_3(X,T) + \mathcal{O}(\varepsilon^4)], 
\end{cases}	
\end{equation*}
with $X = \varepsilon x$ and $T = \varepsilon^2 t$. 
By the standard analysis at the formal order of $\mathcal{O}(\varepsilon^3)$, 
there exists a solution for $(u_3,v_3)$ only if $A_{\pm}(X,T)$ satisfies the NLS equations
\begin{equation}
	\label{NLS}
	{\rm i} \partial_T A_{\pm} = \pm \frac{1}{2} \partial_X^2 A_{\pm} + |A_{\pm}|^2 A_{\pm}.
\end{equation}
The NLS equation (\ref{NLS}) for $A_+$ is focusing and admits bright solitons on the zero background. The NLS equation (\ref{NLS}) for $A_-$ is defocusing and admits dark solitons on the nonzero background. The nonzero background is modulationally unstable for the focusing NLS equation and modulationally (and spectrally) stable for the defocusing NLS equation.

The bright solitons on the zero background exist for $\omega$ in the gap $(-1,1)$ of the purely continuous spectrum of the Dirac operator (\ref{Dirac}) and admit the exponentially decaying tails. The limit $\omega \to -1$ is nontrivial and results in the algebraically decaying tails of the bright soliton. The spectral stability of exponentially decaying bright solitons for $\omega \in (-1,1)$ was established in \cite{KaupLakoba1996}. Their orbital stability was proved for the perturbations in $H^1(\mathbb{R})$ \cite{PS14} and in $L^2(\mathbb{R})$ \cite{CPS16}. The inverse scattering transform in weighted Sobolev spaces of higher regularity was developed in \cite{PS19}, see \cite{Kawata,Lee,Villaroel} for earlier works. The proof of the asymptotic stability of bright solitons and the soliton resolution conjecture was completed in \cite{Cheng23}. 
For the algebraically decaying bright soliton at $\omega = -1$, the orbital 
stability proof is an open problem but the recent studies of 
higher-order rational 
solutions of the MTM system (\ref{MTM}) in  \cite{Han-2024,LiPelin2025,ZhaoPelin2026} (see also related work \cite{Wang1}) suggest their stability in the nonlinear dynamics of soliton interactions.

Compared to the comprehensive studies of bright solitons, dark solitons have received less attention in the context of the modulational stability of the nonzero background. It was shown in \cite{Chen-SAPM-2023} that an algebraic dark soliton appears as the limiting 
case in the family of the exponential dark solitons. The polynomial Darboux matrices for the MTM were introduced in \cite{Deg2015} and were subsequently applied in \cite{Degasperis} to construct the fundamental rogue wave. Higher-order rogue waves were constructed in \cite{He} and \cite{Ye}. Finally, higher-order rational solutions for rogue waves in the MTM were systematically derived 
from solutions of the two-component KP hierarchy in \cite{Baofeng1}, see also \cite{YY-book}. Spectral stability and instability of periodic waves, which include both the nonzero background and the dark solitons on the nonzero background in the limiting cases, were studied in \cite{Cui2025}.

It is well-known from \cite{Degasperis,Ye} that rogue waves on the nonzero background appear when the background is modulationally unstable. On the other hand, dark solitons may exist even if the nonzero background is spectrally unstable, in which case they pick up the spectral instability of the background. The interplay between the spectral stability or instability and the existence regions of the algebraically decaying dark solitons versus the rogue waves has not been studied comprehensively for the MTM system (\ref{MTM}). 
For example, action-angle variables were constructed in \cite{Tian2026} for a particular nonzero background, which our analysis identifies as spectrally unstable. The role of action-angle variables for the nonlinear dynamics of the unstable background is not clear. However, there exists a spectrally stable background, for which the action-angle variables are important for the nonlinear dynamics of dark solitons.

{\em The purpose of this work is to clarify the existence regions for the algebraic dark solitons versus the rogue waves in relation to the spectral stability or instability of the nonzero background in the MTM system (\ref{MTM}).}

\subsection{Main results}

Let us first introduce the nonzero background. We are looking for the plane-wave solution of the MTM system (\ref{MTM}) in the form:
\begin{equation}\label{0-slo}
\begin{cases}
u(x,t) = \rho_{1}e^{{\rm i}(\ell x- \omega t)},\\
v(x,t) = \rho_{2} e^{{\rm i} (\ell x-\omega t)},
\end{cases}
\end{equation}
where $\ell$, $\omega$, $\rho_{1}$, $\rho_{2}$ are real parameters 
and $\rho_1$,$\rho_2$ are nonzero. Substituting (\ref{0-slo}) into (\ref{MTM}), we obtain the linear system
\begin{equation*}
\begin{cases}
\omega-\ell=\rho^2_{2}-\frac{\rho_{2}}{\rho_{1}},\\ \omega+\ell=\rho^2_{1}-\frac{\rho_{1}}{\rho_{2}},
\end{cases}
\end{equation*}
which is uniquely solved for $\ell$ and $\omega$ in terms of $\rho_1$ and $\rho_2$ by 
\begin{equation}
\label{relation}
\begin{cases}
\ell=\frac{1}{2}(\rho^{2}_{1}-\rho^{2}_{2})\left(1-\frac{1}{\rho_{1}\rho_2{}}\right),\\
\omega=\frac{1}{2}(\rho^2_{1}+\rho^2_{2})\left(1-\frac{1}{\rho_{1}\rho_{2}}\right).
\end{cases}
\end{equation}
Two particular cases lead to the spatially homogeneous nonzero background:
\begin{itemize}
	\item $\rho_2 = \frac{1}{\rho_1}$, for which $\ell=\omega=0$,
	\item $\rho_2 = \pm \rho_1$, for which $\ell = 0$ and $\omega = \rho_1^2 \mp 1$.
\end{itemize}

Stability of the nonzero background is considered in the spectral sense. 
By using perturbations defined by 
\begin{equation}\label{disturbance}
\begin{cases}
u(x,t) = \rho_{1}e^{{\rm i}(\ell x- \omega t)}(1+U(x,t)),\\
v(x,t) = \rho_{2} e^{{\rm i} (\ell x-\omega t)}(1+V(x,t)),
\end{cases}
\end{equation}
we can linearize equations of motion and 
expand $(U,V)(\cdot,t) \in L^2(\mathbb{R}) \times L^2(\mathbb{R})$ 
with the Fourier transform 
\begin{equation}\label{FT}
\begin{cases}
U(x,t) = \int_{\mathbb{R}} \left[ \hat{U}_+(k) e^{t \lambda(k)} + 
\hat{\bar{U}}_-(-k) e^{t \bar{\lambda}(-k)} \right] e^{{\rm i} k x} dk,\\
V(x,t) = \int_{\mathbb{R}} \left[ \hat{V}_+(k) e^{t \lambda(k)} + 
\hat{\bar{V}}_-(-k) e^{t \bar{\lambda}(-k)} \right] e^{{\rm i} k x} dk,
\end{cases}
\end{equation}
where $(\hat{U}_+(k),\hat{U}_-(k),\hat{V}_+(k),\hat{V}_-(k))$ are defined 
from a linear homogeneous system for each $k \in \mathbb{R}$. 
There exist four admissible values for $\lambda = \lambda(k)$ for each
$k \in \mathbb{R}$, which are symmetric relative to ${\rm i} \mathbb{R}$: if $\lambda(k)$ is a root, so is $-\bar{\lambda}(k)$. In addition, 
the admissible values are symmetric relative to $\mathbb{R}$ by reflection of $k$: if $\lambda(k)$ is a root, so is $-\lambda(-k)$.

\begin{definition}
	\label{def-stability}
We say that the nonzero background in (\ref{0-slo}) is spectrally stable if 
all admissible solutions in the Fourier form (\ref{FT}) with $\lambda = \lambda(k)$ satisfy $\lambda(k) \in {\rm i} \mathbb{R}$ for every $k \in \mathbb{R}$. This guarantees that the perturbations $(U,V)(\cdot,t) \in L^2(\mathbb{R}) \times L^2(\mathbb{R})$ do not grow exponentially as $t \to +\infty$. Otherwise, the nonzero background in (\ref{0-slo}) is spectrally unstable. 
\end{definition}

Among the spectral instabilities, we classify three distinct cases:
\begin{itemize}
	\item {\bf Modulational instability:} $\exists k_0 > 0 : $ 
	${\rm Re} \lambda(k) > 0$ for $|k| \in (0,k_0)$ for two of the four admissible values of $\lambda(k)$, whereas ${\rm Re} \lambda(k) = 0$ for $k = 0$.
	\item {\bf Side-band instability:} $\exists k_+ > k_- > 0 : $ 
	${\rm Re} \lambda(k) > 0$ for $|k| \in (k_-,k_+)$ for two of the four admissible values of $\lambda(k)$.
	\item {\bf Full-band instability:} $\exists k_0 > 0 : $ 
	${\rm Re} \lambda(k) > 0$ for $k \in (-k_0,k_0)$ for two of the four admissible values of $\lambda(k)$.
\end{itemize}
In the case of modulational instability, there may exist an additional side-band instability with $k_+ > k_- > k_0$ such that ${\rm Re} \lambda(k) > 0$ for $|k| \in (0,k_0) \cup (k_-,k_+)$ for two of the four admissible values of $\lambda(k)$, whereas ${\rm Re} \lambda(k) = 0$ for $|k| \in \{0\} \cup [k_0,k_-] \cup [k_+,\infty)$.

\begin{figure}[htb!]
	\centering
	\includegraphics[width=12cm,height=10cm]{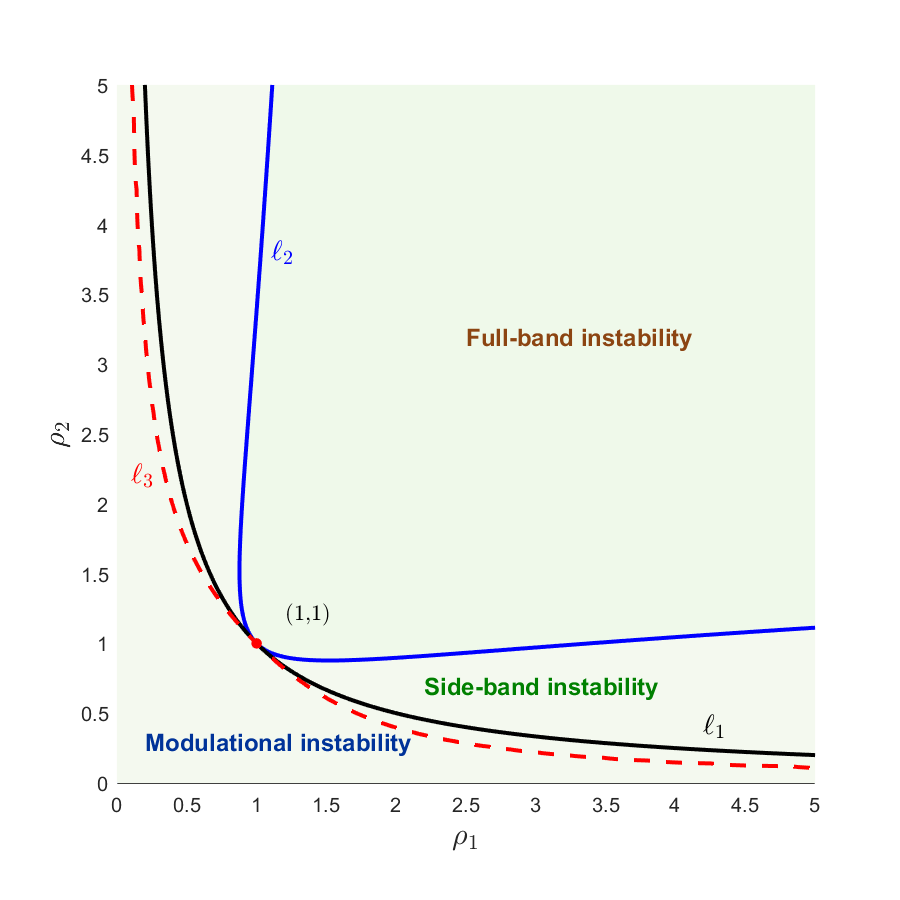}
	\caption{Instability regions in the first quadrant of the $(\rho_1, \rho_2)$ plane. The black ($\ell_1$) and blue ($\ell_2$) curves are given by  $\rho_{1}\rho_{2}=1$ and $4\rho_1^3 \rho_2^3 = (\rho_1^2 + \rho_2^2)^2$ respectively. The red dashed curve ($\ell_3$) is given by $\frac{(\rho^2_{1}-\rho^2_{2})^2}{4 \rho^2_{1}\rho^2_{2}}=\frac{(1-\rho_{1}\rho_{2}) (1+\rho_{1}\rho_{2})^3}{(2\rho_{1}\rho_{2}-1)^3}$. The red dot marks the intersection point $(\rho_1, \rho_2) = (1,1)$ of all curves.}
	\label{fig-stab}
\end{figure}

The  following theorem describes the main result on the stability and instability of the nonzero background. Regions of the first quadrant of the $(\rho_1,\rho_2)$ plane for different unstable cases are shown in Figure \ref{fig-stab}.

\begin{theorem}
	\label{theorem-stab}
	The nonzero background (\ref{0-slo}) is spectrally stable if $\rho_1 \rho_2 < 0$ and is spectrally unstable if $\rho_1 \rho_2 > 0$. Moreover, it exhibits modulational instability if $0 < \rho_1 \rho_2 < 1$, full-band instability if 
	$4 \rho_1^3 \rho_2^3 > (\rho_1^2 + \rho_2^2)^2$, and side-band instability if $\rho_1 \rho_2 > 1$ and $4 \rho_1^3 \rho_2^3 < (\rho_1^2 + \rho_2^2)^2$. The modulational instability is accompanied by an additional side-band instability if $\frac{1}{2} < \rho_1 \rho_2 < 1$ and 
	\begin{align*}
		\frac{(\rho^2_{1}-\rho^2_{2})^2}{4 \rho^2_{1}\rho^2_{2}} > \frac{(1-\rho_{1}\rho_{2}) (1+\rho_{1}\rho_{2})^3}{(2\rho_{1}\rho_{2}-1)^3}.
	\end{align*}
\end{theorem}

\begin{remark}
The MTM system considered in \cite{Tian2026} is related to our system (\ref{MTM}) by the transformation $x \mapsto -x$, $t \mapsto -t$, and $(u,v) \mapsto (u,-v)$. The boundary conditions in \cite[Eq. (2)]{Tian2026} are prescribed as $(u,v) \to (1,-1) e^{{\rm i} \theta_{\pm}}$ as $x \to \pm \infty$ with $\theta_{\pm} \in \mathbb{R}$. Under the transformation, these conditions correspond to the case $\rho_1 = \rho_2 = 1$ in Theorem \ref{theorem-stab}, for which the background is spectrally unstable. It is more interesting to extend the action-angle variables in \cite{Tian2026} to the spectrally stable background with $\rho_1 \rho_2 < 0$, e.g., for $\rho_1 = -\rho_2 = 1$.
\end{remark}

\begin{remark}
	The condition $0 < \rho_1 \rho_2 < 1$ for the modulational instability of the nonzero background was previously obtained in \cite{Degasperis}. However, the complete classification of the side-band and full-band instabilities in Theorem \ref{theorem-stab} is a new result.
\end{remark}

Next, we consider algebraically decaying dark solitons and rogue waves 
on the nontrivial background (\ref{0-slo}). Based on the exact expressions 
from \cite{Chen-SAPM-2023,Baofeng1}, we prove the following theorem.

\begin{theorem}
	\label{theorem-soliton}
If either $\rho_1\rho_2 < 0$ or $\rho_1 \rho_2 > 1$ and $(\rho_1^2 + \rho_2^2)^2 \neq 4 \rho_1^3 \rho_2^3$, there exist exactly two 
algebraically decaying dark solitons among the traveling wave solutions 
of the MTM system (\ref{MTM}) in the form:
\begin{equation}
\label{TW-dark}
u(x,t) = \rho_{1}e^{{\rm i}(\ell x- \omega t)} \mathfrak{u}_{\pm}(x-ct),\qquad
v(x,t) = \rho_{2} e^{{\rm i} (\ell x-\omega t)} \mathfrak{v}_{\pm}(x-ct),
\end{equation}
where $\mathfrak{u}_{\pm},\mathfrak{v}_{\pm}$ are rational functions satisfying 
$|\mathfrak{u}_{\pm}- 1|, |\mathfrak{v}_{\pm} - 1| = \mathcal{O}(|x|^{-1})$ as $|x| \to \infty$ with the maximal normalized total intensity given by 
\begin{equation}
\label{max-elevation-dark}
\max_{x \in \mathbb{R}} |\mathfrak{u}_{\pm}(x)|^2 + |\mathfrak{v}_{\pm}(x)|^2 = 2 + 8 \frac{\rho_1 \rho_2 - 1}{\rho_1 \rho_2}, 
\end{equation}
and with two distinct values of the wave speed $c = c_{\pm}$ given by  
\begin{equation}
\label{speeds}
c_{\pm} = \frac{\rho_{1}^{4} - \rho_{2}^{4} \pm 4 \rho_{1}^{2} \rho_{2}^{2} \sqrt{\rho_1 \rho_2 (\rho_1 \rho_2 - 1)}}{(\rho_{1}^{2}+\rho_{2}^{2})^{2} - 4 \rho_{1}^{3} \rho_{2}^{3}}.
\end{equation}
The values of $c_{\pm}$ in (\ref{speeds}) are the end points 
in the existence region of exponentially decaying dark solitons 
which exist only if 
\begin{itemize}
	\item $c \in (c_-,c_+)$ for either  $\rho_1 \rho_2 < 0$ or $\rho_1 \rho_2 > 1$ and $4 \rho_1^3 \rho_2^3 < (\rho_1^2 + \rho_2^2)^2$,
\end{itemize}
or
\begin{itemize}
	\item $c \in (-\infty,c_+) \cup (c_-,\infty)$ for $4 \rho_1^3 \rho_2^3 > (\rho_1^2 + \rho_2^2)^2$.
\end{itemize}
No algebraically decaying dark solitons exist for $0 < \rho_1 \rho_2 < 1$, for which the MTM system (\ref{MTM}) admits only one fundamental rogue wave solution in the form 
\begin{equation}
\label{rogue}
u(x,t) = \rho_{1}e^{{\rm i}(\ell x- \omega t)} \hat{u}(x,t),\qquad
v(x,t) = \rho_{2} e^{{\rm i} (\ell x-\omega t)} \hat{v}(x,t),
\end{equation}
where $\hat{u},\hat{v}$ are rational functions satisfying 
$|\hat{u} - 1|, |\hat{v} - 1| = \mathcal{O}((|x| + |t|)^{-1})$ as $|x| + |t| \to \infty$ with the maximal normalized total intensity given by 
\begin{equation}
\label{max-elevation-rogue}
\max_{(x,t) \in \mathbb{R}^2} |\hat{u}(x,t)|^2 + |\hat{v}(x,t)|^2 = 18.
\end{equation}
\end{theorem}

\begin{remark}
	The solutions (\ref{TW-dark}) and (\ref{rogue}) are unique modulo the translational and rotational symmetries of the MTM system (\ref{MTM}). It is remarkable that the maximal normalized total intensity of the algebraic dark solitons in  (\ref{max-elevation-dark}) is identical between the two solitons and that the maximal normalized total intensity of the rogue wave in (\ref{max-elevation-rogue}) is independent of the parameters $\rho_1$ and $\rho_2$.
\end{remark}

\begin{remark}
	Explicit expressions for two algebraic dark solitons (\ref{TW-dark}) and the rogue wave (\ref{rogue}) in Theorem \ref{theorem-soliton} are presented in Section \ref{sec-dark} together with the solution surfaces. Compared to \cite{Chen-SAPM-2023}, where only one algebraic dark soliton was obtained, we construct two different types of the algebraic dark solitons and clarify the spectral stability of their background if $\rho_1 \rho_2 < 0$ and their spectral instability if $\rho_1 \rho_2 > 1$, due to Theorem \ref{theorem-stab}. 
\end{remark}

\begin{remark}
	The exceptional case $(\rho_1^2 + \rho_2^2)^2 = 4 \rho_1^3 \rho_2^3$ for $\rho_1 \rho_2 > 1$ corresponds to the curve $\ell_2$ in Figure \ref{fig-stab}, where the side-band instability region for $(\rho_1^2 + \rho_2^2)^2 > 4 \rho_1^3 \rho_2^3$ is connected to the full-band instability region for $(\rho_1^2 + \rho_2^2)^2 < 4 \rho_1^3 \rho_2^3$. There exists only one algebraically decaying dark soliton with the wave speed given by 
\begin{equation}
\label{speed-unique}
c=\frac{(\rho^2_{1}-\rho^2_{2})^2+4\rho^3_{1} \rho^3_{2}}{2 (\rho^2_{1}+\rho^2_{2})  (\rho^2_{1}-\rho^2_{2}) }.
\end{equation}
\end{remark}

\begin{remark}
	The condition $0 < \rho_1 \rho_2 < 1$ for the existence of rogue waves was found in all previous works on rogue waves: \cite{Baofeng1,Degasperis,He,Ye}. Higher-order rogue waves, which generalize (\ref{rogue}) with the higher-order rational functions $\hat{u},\hat{v}$, were discussed in \cite{He} and systematically derived in \cite{Baofeng1}. Compared to \cite{Baofeng1} and in agreement with \cite{Degasperis}, we clarify that the condition $0 < \rho_1 \rho_2 < 1$ for the rogue wave existence coincides with the criterion for the modulational instability of the rogue wave background.
\end{remark}

\begin{remark}
It is an open problem for further studies to construct a hierarchy of higher-order algebraic dark solitons, if it exists, which would 
satisfy 
\begin{equation*}
u(x,t) = \rho_{1}e^{{\rm i}(\ell x- \omega t)} \mathfrak{u}(x-ct,t),\qquad
v(x,t) = \rho_{2} e^{{\rm i} (\ell x-\omega t)} \mathfrak{v}(x-ct,t),
\end{equation*}
with the higher-order rational functions $\mathfrak{u},\mathfrak{v}$ satisfying  
$|\mathfrak{u} - 1|, |\mathfrak{v} - 1| = \mathcal{O}(|x-ct|^{-1})$ as $|x| \to \infty$ for any $t \in \mathbb{R}$. A similar hierarchy of higher-order algebraic bright solitons on the zero background was recently discovered in \cite{Han-2024,ZhaoPelin2026}. 
\end{remark}

Theorem \ref{theorem-stab} is proven in Section \ref{sec-stab}. Theorem \ref{theorem-soliton} is proven in Section \ref{sec-dark}. The paper is concluded with Section \ref{sec-conclusion}.

\section{Stability of the nonzero background}
\label{sec-stab}

\subsection{Linearized equations of motion}

To prove Theorem \ref{theorem-stab}, we first obtain the linearized equations of motion. Substituting (\ref{disturbance}) into (\ref{MTM}) and using \eqref{relation} to eliminate $\ell$ and $\omega$, we truncate the quadratic and cubic terms for $(U,V)$ and obtain 
\begin{equation}
\label{MTM1}
	\begin{cases}
		{\rm i} (U_{t}+U_{x})=\frac{\rho_{2}}{\rho_{1}}U+
		\frac{\rho_{2}}{\rho_{1}}(\rho_{1}\rho_{2}-1)V+\rho^2_{2}\bar{V},\\
		{\rm i}(V_{t}-V_{x})=\frac{\rho_{1}}{\rho_{2}}(\rho_{1}\rho_{2}-1)U+\frac{\rho_{1}}{\rho_{2}}V+\rho^2_{1}\bar{U}.
	\end{cases}
\end{equation}
By using the Fourier transform (\ref{FT}) for solutions 
of the system (\ref{MTM1}) and the complex-conjugate system, we obtain a linear homogeneous 
system for $(\hat{U}_+(k), \hat{U}_-(k), \hat{V}_+(k),\hat{V}_-(k))$ 
with the spectral parameter $\lambda(k)$ at each 
Fourier mode with wave number $k \in \mathbb{R}$. Dropping the argument of 
$k \in \mathbb{R}$, we write the linear system as follows:
\begin{equation*}
	\begin{cases}
		({\rm i} \lambda - \frac{\rho_{2}}{\rho_{1}}-k)\hat{U}_{+} - \frac{\rho_{2}}{\rho_{1}}(\rho_{1}\rho_{2}-1)\hat{V}_{+} - \rho^{2}_{2}\hat{V}_{-}=0,\\
		(-{\rm i} \lambda - \frac{\rho_{2}}{\rho_{1}}+k)\hat{U}_{-} - \frac{\rho_{2}}{\rho_{1}}(\rho_{1}\rho_{2}-1)\hat{V}_{-} - \rho^{2}_{2}\hat{V}_{+}=0,\\
		({\rm i} \lambda - \frac{\rho_{1}}{\rho_{2}}+k)\hat{V}_{+} - \frac{\rho_{1}}{\rho_{2}}(\rho_{1}\rho_{2}-1)\hat{U}_{+} - \rho^{2}_{1}\hat{U}_{-}=0,\\
		(-{\rm i} \lambda - \frac{\rho_{1}}{\rho_{2}}-k)\hat{V}_{-} - \frac{\rho_{1}}{\rho_{2}}(\rho_{1}\rho_{2}-1)\hat{U}_{-} - \rho^{2}_{1}\hat{U}_{+}=0.
	\end{cases}
\end{equation*}
There exists a nonzero solution for $(\hat{U}_+, \hat{U}_-, \hat{V}_+,\hat{V}_-)$ if and only if $\lambda$ is a root of the characteristic equation given by
\begin{align*} 
	D(\lambda) = \begin{vmatrix}
		{\rm i} \lambda - k - \frac{\rho_2}{\rho_1} & 0 & -\frac{\rho_2}{\rho_1} (\rho_1 \rho_2 - 1) & -\rho_2^2 \\
		0 & -{\rm i} \lambda + k - \frac{\rho_2}{\rho_1} & -\rho_2^2 & -\frac{\rho_2}{\rho_1} (\rho_1 \rho_2 - 1) \\
		-\frac{\rho_1}{\rho_2} (\rho_1 \rho_2 - 1) & -\rho_1^2 & {\rm i} \lambda + k - \frac{\rho_1}{\rho_2} & 0 \\
		-\rho_1^2 & -\frac{\rho_1}{\rho_2} (\rho_1 \rho_2 - 1) & 0 & -{\rm i} \lambda - k - \frac{\rho_1}{\rho_2}                   
	\end{vmatrix}.
\end{align*}
Expanding the determinant, we obtain a quartic polynomial in $\lambda$:
\begin{align}
	D(\lambda) &= 
\lambda^4 + \lambda^2 \left(2k^2-4\rho_{1}\rho_{2}+\frac{(\rho^{2}_{2}+\rho^2_{1})^2}{\rho^{2}_{1}\rho^2_{2}}\right) + 2 {\rm i} \lambda k \frac{\rho^{4}_{1}-\rho^{4}_{2}}{\rho^{2}_{1}\rho^{2}_{2}}  \notag \\
& \quad + k^2\left(k^2-4\rho_{1}\rho_{2}-\frac{(\rho^2_{1}-\rho^2_{2})^2}{\rho^2_{1}\rho^2_{2}}\right).
\label{characteristic2}	
\end{align}
There exist four admissible roots $\lambda$ of the quartic polynomial $D(\lambda)$. It follows from the symmetries of coefficients of $D(\lambda)$  that if $\lambda = \lambda(k)$ is a root, so are $-\bar{\lambda}(k)$, $-\lambda(-k)$, and $\bar{\lambda}(-k)$. For further use, we rewrite the polynomial 
(\ref{characteristic2}) for $\lambda={\rm i}\Omega$ as $P(\Omega) := D({\rm i} \Omega)$ with 
\begin{align}
P(\Omega) &= \Omega^4 - \Omega^2 \left(2k^2-4\rho_{1}\rho_{2}+\frac{(\rho^{2}_{2}+\rho^2_{1})^2}{\rho^{2}_{1}\rho^2_{2}}\right) - 2 \Omega k \frac{\rho^{4}_{1}-\rho^4_{2}}{\rho^2_{1}\rho^2_{2}} \notag \\ & \quad + k^2 \left(k^2-4\rho_{1}\rho_{2}-\frac{(\rho^2_{1}-\rho^2_{2})^2}{\rho^2_{1}\rho^2_{2}}\right).
	\label{characteristic3}	
\end{align}
According to Definition \ref{def-stability}, the nonzero background (\ref{0-slo}) is spectrally stable if all four roots $\lambda$ of $D(\lambda)$ are on ${\rm i} \mathbb{R}$ for every $k \in \mathbb{R}$. Equivalently, it is spectrally stable if all four roots $\Omega$ of $P(\Omega)$ are on $\mathbb{R}$ for every $k \in \mathbb{R}$. The proof of Theorem \ref{theorem-stab} is obtained 
from analysis of the polynomial $P(\Omega)$ based on Propositions \ref{prop-stab} and \ref{prop-instab}.

\subsection{Spectral stability for $\rho_1 \rho_2 < 0$} 

\begin{prop}
	\label{prop-stab}
If $\rho_1 \rho_2 < 0$, all four roots $\Omega$ of $P(\Omega)$ belong to $\mathbb{R}$ for every $k\in\mathbb{R}$. 
\end{prop}

\begin{proof}
We analyze $P(\Omega)$ given by (\ref{characteristic3}) for every $k \in \mathbb{R}$ if $\rho_1 \rho_2 < 0$. For convenience, we write
\begin{equation}
\label{alpha-1-2}
	\alpha_1:=\frac{\rho_1^2+\rho_2^2}{2\rho_1\rho_2},
	\qquad
	\alpha_2:=\frac{\rho_1^2-\rho_2^2}{2\rho_1\rho_2},
\end{equation}
and represent $P(\Omega)$ in the equivalent form:
	\begin{equation*}
		P(\Omega) =	(\Omega^2-k^2+2\rho_1 \rho_2)^2 - 4(\alpha_1\Omega+\alpha_2k)^2-4\rho^2_1 \rho^2_2.
	\end{equation*}
Therefore, $\Omega$ is a root of $P(\Omega)$ if and only if either $F_+(\Omega) = 0$ or $F_-(\Omega) = 0$, where 
	\begin{align}		
	F_+(\Omega) :=\Omega^2-k^2+2 \rho_1 \rho_2-2\sqrt{(\alpha_1\Omega+\alpha_2k)^2+\rho^2_1 \rho^2_2} \label{F-plus}
	\end{align}
	and
	\begin{align}
	F_-(\Omega) :=-\Omega^2+k^2-2\rho_1 \rho_2-2\sqrt{(\alpha_1\Omega+\alpha_2k)^2+\rho^2_1 \rho^2_2}	\label{F-minus}
	\end{align}
are used in the factorization: $P(\Omega) = -F_+(\Omega) F_-(\Omega)$. 
Since $(\alpha_1\Omega+\alpha_2k)^2+\rho^2_1 \rho^2_2 > 0$ for every $\Omega \in \mathbb{R}$ and $k \in \mathbb{R}$, the functions $F_{\pm}(\Omega) : \mathbb{R} \to \mathbb{R}$ are smooth with respect to both $\Omega \in \mathbb{R}$ and $k \in \mathbb{R}$. Let us define for every $k \in \mathbb{R}$, 
	\[
	\Omega_*:=-\frac{\alpha_2 }{\alpha_1}k.
	\]
Since $\alpha_1\Omega_*+\alpha_2 k=0$ and 
$2 \sqrt{(\alpha_1\Omega_{*}+\alpha_2k)^2+\rho^2_1 \rho^2_2}=-2\rho_1\rho_2 > 0$, 
we obtain 
	\begin{align*}
		F_+(\Omega_*) &= \left[\left(\frac{\rho^2_1-\rho^2_2}{\rho^2_1+\rho^2_2}\right)^2-1\right]k^2+4\rho_1 \rho_2 = -\frac{4 \rho^2_1 \rho^2_2 k^2}{(\rho^2_1+\rho^2_2)^2} + 4\rho_1\rho_2,  \\
		F_-(\Omega_*) &=  \left[1-\left(\frac{\rho^2_1-\rho^2_2}{\rho^2_1+\rho^2_2}\right)^2\right] k^2 = \frac{4 \rho^2_1 \rho^2_2 k^2}{(\rho^2_1+\rho^2_2)^2}.
	\end{align*}
On the other hand, as $|\Omega|\to\infty$, we have 
	\[
	F_+(\Omega) \to +\infty,
	\qquad
	F_-(\Omega) \to -\infty,
	\]
because $2\sqrt{(\alpha_1\Omega+\alpha_2k)^2+\rho^2_1 \rho^2_2}=\mathcal{O}(|\Omega|)$ as $|\Omega| \to \infty$ while the quadratic terms dominate.
	
If $k \neq 0$, then $F_+(\Omega_*) < 0$ and $F_-(\Omega_*) > 0$. By the intermediate value theorem, each of $F_+$ and $F_-$ has at least one zero in $(-\infty,\Omega_*)$ and at least one zero in $(\Omega_*,\infty)$. Thus,  $P(\Omega) : \mathbb{R}\to \mathbb{R}$ has at least four real roots. Since $P(\Omega)$ is a quartic polynomial in $\Omega$, it has at most four roots. Therefore, it has exactly four real roots for $k \neq 0$.

If $k=0$, then roots of $P(\Omega)$ can be found explicitly from 
\begin{align}
\label{P-0}
P(\Omega) |_{k = 0} = \Omega^2
\left[ \Omega^2 - \left( -4\rho_1\rho_2 + \frac{(\rho_1^2+\rho_2^2)^2}{\rho_1^2\rho_2^2} \right) \right],
\end{align}
where the coefficient inside the bracket is strictly positive. Hence,  $P(\Omega)$ has a double zero root and two simple real roots. 
\end{proof}

\subsection{Spectral instability for $\rho_1 \rho_2 > 0$} 

We first obtain three technical results in the following lemmas.

\begin{lemma}
	\label{lem 1}
Assume that $\rho_1 \rho_2 > 0$ and define
	\begin{align*}
	M := 4\rho_1^3\rho_2^3 - (\rho_1^2+\rho_2^2)^2.
	\end{align*}
In addition to the double zero root, $P(\Omega) |_{k=0}$ has two simple real roots if $M < 0$ and two simple purely imaginary roots if $M > 0$.
\end{lemma}

\begin{proof}
If $k=0$, then we use (\ref{P-0}) and rewrite it as 
\begin{align*}
P(\Omega) |_{k = 0} = \Omega^2
\left[ \Omega^2 + \left( 4\rho_1\rho_2 - \frac{(\rho_1^2+\rho_2^2)^2}{\rho_1^2\rho_2^2} \right) \right] = \Omega^2 \left( \Omega^2 + \frac{M}{\rho_1^2\rho_2^2} \right),
\end{align*}
from which the assertion holds.
\end{proof}

\begin{lemma}\label{lem 2}
	If $\rho_1 \rho_2 > 0$ and $M < 0$ in Lemma \ref{lem 1}, 
	then there is $k_0 > 0$ such that for every $|k| \in (0,k_0)$, $P(\Omega)$ has four simple real roots if $\rho_1 \rho_2 > 1$ and two simple real roots and two simple complex-conjugate roots if $\rho_1 \rho_2 < 1$. 
\end{lemma}

\begin{proof}
	We use $\alpha_1$, $\alpha_2$ in (\ref{alpha-1-2}) and introduce $F_+(\Omega)$, $F_-(\Omega)$ in (\ref{F-plus}), (\ref{F-minus}) so that 
	$$
	P(\Omega) = -F_+(\Omega) F_-(\Omega).
	$$ 
	If $\rho_1 \rho_2 > 0$, then $F_+(0) |_{k=0} = 0$ and $F_{-}(0) |_{k=0} = -4 \rho_{1}\rho_{2} < 0$. Since $F_{\pm}(\Omega) : \mathbb{R}\to \mathbb{R}$ are smooth with respect to $(\Omega,k)$ near $(0,0)$, the double zero root of $P(\Omega) |_{k=0}$ in Lemma \ref{lem 1} coincides with the double zero root of $F_+(\Omega) |_{k=0}$. 

In addition, since $\sqrt{\alpha^2_{1} \Omega^2+\rho^2_{1}\rho^2_{2}} \geq \rho_{1}\rho_{2} >0$, we obtain 
\begin{align*}
F_{-}(\Omega) |_{k=0} &=-\Omega^2-2\rho_1 \rho_2-2\sqrt{\alpha^2_{1} \Omega^2+\rho^2_{1}\rho^2_{2}}\\
&\leq -\Omega^2-4\rho_{1}\rho_{2},
\end{align*}
which implies that $F_{-}(\Omega) |_{k=0} < 0$ for every $\Omega \in \mathbb{R}$ 
and does not have any real roots. If $M < 0$, then the two simple real nonzero roots of $P(\Omega) |_{k=0}$ in Lemma \ref{lem 1} are roots of $F_+(\Omega) |_{k=0}$ away from the origin. 

It remains to study how the double zero root of $F_+(\Omega) |_{k=0}$ splits for small $k \neq 0$. To do so, we expand the smooth function $F_{+}$ in $(\Omega,k)$ near $(0,0)$ in the power series:
	\begin{align*}
	F_{+}(\Omega) &=\Omega^2-k^2+2\rho_{1}\rho_{2}-2 \rho_{1}\rho_{2}\left[ 1+\frac{1}{2} \frac{(\alpha_{1} \Omega+\alpha_{2}k)^2}{\rho^2_{1}\rho^2_{2}} +\mathcal{O}((\alpha_{1} \Omega+\alpha_{2}k)^4) \right]\\
	&=\Omega^2\left(1-\frac{\alpha^2_{1}}{\rho_{1}\rho_{2}}\right)-\frac{2 \alpha_{1} \alpha_{2}}{\rho_{1}\rho_{2}}\Omega k -k^2\left(1+\frac{\alpha^2_{2}}{\rho_{1}\rho_{2}}\right)+\mathcal{O}((\alpha_{1} \Omega+\alpha_{2}k)^4),
	\end{align*}
where 
	\begin{align*}
	1-\frac{\alpha^2_{1}}{\rho_{1}\rho_{2}}=\frac{4\rho^3_{1}\rho^3_{2}-(\rho^2_{1}+\rho^2_{2})^2}{4 \rho^3_{1}\rho^3_{2}} \neq 0
	\end{align*}
	by the assumption $M < 0$, which is equivalent to $4\rho^3_{1}\rho^3_{2}-(\rho^2_{1}+\rho^2_{2})^2<0$. 
	By the implicit function theorem, for every small $k$, there exist two small roots $\Omega = \Omega_{\pm}(k)$ of $F_{+}(\Omega)$ which can be 
	expanded as 
	\begin{align}
	\label{def-beta}
	\Omega_{\pm}(k)=\beta_{\pm}k+\mathcal{O}(k^2), \quad \mbox{\rm as} \;\; k \to 0,
	\end{align}
with two values for $\beta_{\pm}$ found from the quadratic equation
	\begin{align}
	\label{quad-beta}
	\left(1-\frac{\alpha^2_{1}}{\rho_{1}\rho_{2}}\right)\beta^2-2\frac{\alpha_{1}\alpha_{2}}{\rho_{1}\rho_{2}} \beta-\left(1+\frac{\alpha^2_2}{\rho_{1}\rho_{2}}\right)=0.
	\end{align}
The discriminant of the quadratic equation (\ref{quad-beta}) is given by 
	\begin{align*}
	\Delta = \left(\frac{2 \alpha_{1} \alpha_{2}}{\rho_{1}\rho_{2}}\right)^2 + 4 \left(1-\frac{\alpha^2_{1}}{\rho_{1}\rho_{2}}\right)\left(1+\frac{\alpha^2_2}{\rho_{1}\rho_{2}}\right)=4\left(1-\frac{1}{\rho_{1}\rho_{2}}\right),
	\end{align*}
where we have used 
\begin{equation*}
\alpha_1^2-\alpha_2^2
=
\frac{(\rho_1^2+\rho_2^2)^2-(\rho_1^2-\rho_2^2)^2}{4\rho_1^2\rho_2^2}
=1.
\end{equation*}

If $\rho_{1}\rho_{2} > 1$, then $\Delta > 0$ and $\beta_{\pm}$ in (\ref{def-beta}) are real distinct numbers so that $\Omega_{\pm}$ are two simple real roots of $F_+(\Omega)$ near $0$ for every small $|k|$.  In addition, $F_+(\Omega)$ has two simple real roots away from $0$ for every small $|k|$. Hence, $P(\Omega)$ has four simple real roots for small $k \neq 0$.

If $\rho_1 \rho_2 < 1$, then $\Delta < 0$ and $\beta_{\pm}$ in (\ref{def-beta}) are complex-conjugate so that $\Omega_{\pm}$ are two simple complex-conjugate roots of $F_+(\Omega)$  near $0$ for every small $|k|$.  In addition, $F_+(\Omega)$ has two simple real roots away from $0$ for every small $|k|$. Hence, $P(\Omega)$ has two simple real roots and two simple complex-conjugate roots for small $k \neq 0$.	
\end{proof}

\begin{lemma}
	\label{lem 3}
If $\Omega_{1} \in \mathbb{R}$ is a root of $F_{-}(\Omega)$ for any fixed $k \in \mathbb{R}$, then $F_{+}(\Omega_{1}) \neq 0$.
\end{lemma}

\begin{proof}
	We argue by contradiction. If $F_-(\Omega_1) = F_+(\Omega_1) = 0$, then 
	\begin{align*}
	\Omega_1^2-k^2+2\rho_1\rho_2 - 
	2\sqrt{(\alpha_1 \Omega_1 + \alpha_2 k)^2 +\rho^2_1 \rho^2_2} &= 0, \\
	\Omega_1^2-k^2+2\rho_1\rho_2 + 
	2\sqrt{(\alpha_1 \Omega_1 + \alpha_2 k)^2+\rho^2_1 \rho^2_2} &=0,
	\end{align*}
which implies $\sqrt{(\alpha_1 \Omega_1 + \alpha_2 k)^2 +\rho^2_1 \rho^2_2} = 0$, a contradiction. Therefore, both $F_+$ and $F_-$ cannot be zero at the same $\Omega_1 \in \mathbb{R}$ for any fixed $k \in \mathbb{R}$.
\end{proof}

Next, we obtain the instability result based on Lemmas \ref{lem 1}, \ref{lem 2}, and \ref{lem 3}.

\begin{prop}
	\label{prop-instab}
	Assume that $\rho_1 \rho_2 > 0$ and consider roots of $P(\Omega)$ given by (\ref{characteristic3}).
	\begin{enumerate}
		\item If $M > 0$, there exists $k_0 > 0$ such that $P(\Omega)$ has four real roots for $|k| > k_0$ and two real and two complex-conjugate roots for $|k| < k_0$. 
		
		\item If $M < 0$ and $\rho_1 \rho_2 > 1$, there exists $k_+ > k_- > 0$ such that $P(\Omega)$ has four real roots for $|k| < k_-$ and $|k| > k_+$ and two real and two complex-conjugate roots for $k_- < |k| < k_+$. 
		
\item If $M < 0$, $\rho_1 \rho_2 < 1$, and $\alpha^2_{2}<\frac{(1-\rho_{1}\rho_{2}) (1+\rho_{1}\rho_{2})^3}{(2\rho_{1}\rho_{2}-1)^3}$, there exists $k_0 > 0$ such that $P(\Omega)$ has four real roots for $|k| > k_0$ and $k = 0$ and two real and two complex-conjugate roots for $0 < |k| < k_0$.
		
\item If $M < 0$, $\frac{1}{2} < \rho_1 \rho_2 < 1$, 
and $\alpha^2_{2} > \frac{(1-\rho_{1}\rho_{2}) (1+\rho_{1}\rho_{2})^3}{(2\rho_{1}\rho_{2}-1)^3}$, there exists $k_+ > k_- > k_0 > 0$ such that $P(\Omega)$ has four real roots for $|k| > k_+$, $k_0 < |k| < k_-$ and $k = 0$ and two real and two complex-conjugate roots for $0 < |k| < k_0$ and $k_- <|k| < k_+$. 
	\end{enumerate}	
\end{prop}

\begin{proof}	
	Since $P(\Omega,k) = -F_+(\Omega) F_-(\Omega)$, we analyze roots of $F_+(\Omega)$ and $F_-(\Omega)$ independently. Recall that $F_{\pm}(\Omega) : \mathbb{R}\to \mathbb{R}$ are smooth with respect to both $\Omega \in \mathbb{R}$ and $k \in \mathbb{R}$. 
	
	Let us define for every $k \in \mathbb{R}$, 
	\[
	\Omega_*:=-\frac{\alpha_2 }{\alpha_1}k.
	\]
Since $\alpha_1\Omega_*+\alpha_2 k=0$ and $\rho_1 \rho_2 > 0$, we obtain 
	\begin{align*}
F_+(\Omega_*) &= \left[\left(\frac{\rho^2_1-\rho^2_2}{\rho^2_1+\rho^2_2}\right)^2-1\right]k^2 = -\frac{4 \rho^2_1 \rho^2_2 k^2}{(\rho^2_1+\rho^2_2)^2}, \\
F_{-}(\Omega_{*}) &= \left[1 - \left(\frac{\rho^2_1-\rho^2_2}{\rho^2_1+\rho^2_2}\right)^2\right] k^2 - 4 \rho_1 \rho_2 = \frac{4 \rho^2_1 \rho^2_2 k^2}{(\rho^2_1 +\rho^2_2)^2} - 4 \rho_1 \rho_2.
	\end{align*}
On the other hand, as $|\Omega| \to \infty$, we have as in Proposition \ref{prop-stab},	
	\[
F_+(\Omega) \to +\infty,
\qquad
F_-(\Omega) \to -\infty.
\]	
Since $F_+(\Omega_*) < 0$ for $k \neq 0$ and $F_+(\Omega) \to +\infty$ as $|\Omega| \to \infty$, there exist at least two real roots of $F_+(\Omega)$ for every $k \neq 0$. Similarly, $F_-(\Omega_*) > 0$ for sufficiently large $|k|$ and $F_-(\Omega) \to -\infty$ as $|\Omega| \to \infty$ so that there exist at least two real roots of $F_-(\Omega)$ for sufficiently large $|k|$. Since $P(\Omega)$ is a quartic polynomial in $\Omega$, it has at most four roots. Thus, there exists $k_1 > 0$ such that $P(\Omega)$ admits four real roots for $|k| > k_1$, two of which 
are roots of $F_+(\Omega)$ and two of which are roots of $F_-(\Omega)$ with 
$k_1$ estimated from $F_-(\Omega_*) > 0$ as
$$
k_1 := \frac{\rho_1^2 + \rho_2^2}{\sqrt{\rho_1\rho_2}}.
$$

Next, we show that there exists $k_2 \in (0,k_1)$ such that $F_-(\Omega)$ does not have any real roots for $|k| < k_2$. To do so, we prove that $F_-(\Omega)$ is a strictly concave function. Indeed, differentiating $F_-$ in $\Omega$, we obtain 
	\begin{align*}
	F'_{-}(\Omega) = -2\Omega - \frac{2 \alpha_{1}(\alpha_{1}\Omega+\alpha_{2}k)}{\sqrt{(\alpha_{1}\Omega+\alpha_{2}k)^2+\rho^2_1 \rho^2_2}}
	\end{align*}
	and
	\begin{align*}
	F^{''}_{-}(\Omega) = -2 - \frac{2 \alpha^2_{1}\rho^2_{1}\rho^2_{2}}{
		(\sqrt{(\alpha_{1}\Omega+\alpha_{2}k)^2+\rho^2_1 \rho^2_2})^3},
	\end{align*}
so that $F''_-(\Omega) < 0$ for every $\Omega \in \mathbb{R}$. Consequently, the graph of $F_-(\Omega)$ can intersect the horizontal axis at most twice, as it does 
for sufficiently large $|k| > k_1$. It remains to show that $F_-(\Omega)$ is strictly negative and does not have any real roots for sufficiently small $|k|$. To do so, we use $\sqrt{(\alpha_1\Omega+\alpha_2k)^2+\rho^2_1 \rho^2_2} \geq |\alpha_1 \Omega+\alpha_2 k|\geq |\alpha_{1} \Omega| - |\alpha_2 k|$ and estimate 
	\begin{align*}
	F_{-}(\Omega) &\leq -\Omega^2+k^2-2\rho_1 \rho_2-2|\alpha_1 \Omega+\alpha_2 k|\\
	&\leq -\Omega^2+k^2-2\rho_1 \rho_2-2 |\alpha_1 \Omega| + 2 |\alpha_2 k|\\
	&\leq k^2 + 2 |\alpha_2 k| - 2 \rho_1 \rho_2.
	\end{align*}
Hence $F_-(\Omega) < 0$ for every $\Omega \in \mathbb{R}$ if $|k| < k_2$, where  
	\begin{align*}
	k_2 :=\sqrt{\alpha^2_{2}+2 \rho_1 \rho_2}-|\alpha_{2}| = \frac{2\rho_{1}\rho_{2}}{\sqrt{\alpha_{2}^{2} + 2\rho_{1}\rho_{2}} + |\alpha_{2}|} > 0.
	\end{align*}
is obtained from the root of $k^2+2 |\alpha_2 k|-2 \rho_1 \rho_2 = 0$. 

Let us now show that $k_2 < k_1$. Since $\sqrt{\alpha_{2}^{2} + 2\rho_{1}\rho_{2}} + |\alpha_{2}| \ge \sqrt{2\rho_{1}\rho_{2}}$, we have 
$$
k_2 = \frac{2\rho_{1}\rho_{2}}{\sqrt{\alpha_{2}^{2} + 2\rho_{1}\rho_{2}} + |\alpha_{2}|} \le \frac{2\rho_{1}\rho_{2}}{\sqrt{2\rho_{1}\rho_{2}}} = \sqrt{2\rho_{1}\rho_{2}}.
$$ 
Furthermore, since $\alpha_1 \ge 1$, it follows that 
$$
k_1 =\frac{\rho^2_{1}+\rho^2_{2}}{\sqrt{\rho_{1}\rho_{2}}}= 2\alpha_{1}\sqrt{\rho_{1}\rho_{2}} \ge 2\sqrt{\rho_{1}\rho_{2}} > \sqrt{2\rho_{1}\rho_{2}}.
$$
Therefore, we have $0 < k_2 < k_1$. 

Let us define
\begin{equation*}
m(k) = \max_{\Omega \in \mathbb{R}} F_{-}(\Omega).
\end{equation*}
Since $F_{-}(\Omega) \to -\infty$ as $|\Omega| \to \infty$ for any $k \in \mathbb{R}$, the maximum value $F_-(\Omega)$ is always attained for any $k \in \mathbb{R}$. Since $F_-(\Omega)$ is a strictly concave function, its maximum value denoted by $m(k)$ is unique. We have $m(k) > 0$  for $|k| > k_1$ and $m(k) < 0$ for $|k| < k_2$. Since $m(k)$ is continuous on $[k_2,k_1]$ the intermediate value theorem guarantees the existence of at least one point $k_0 \in (k_2,k_1)$ such that $m(k_0) = 0$. 

We show that $m'(k) > 0$ for $k > 0$. By monotonicity of $m(k)$ in $k$, the value of $k_0 \in (k_2,k_1)$, for which $m(k_0)=0$, is unique. 
To show that $m'(k) > 0$ for $k > 0$, we denote
the argument of $\max_{\Omega \in \mathbb{R}} F_{-}(\Omega)$ by  $\Omega(k)$ for any $k \in \mathbb{R}$. Then, it satisfies $F_-'(\Omega(k)) = 0$. Since $F_-''(\Omega(k)) \neq 0$, the implicit function theorem ensures that $\Omega(k)$ is continuous and differentiable for any $k \in \mathbb{R}$. Applying the chain rule yields
\begin{align*}
m'(k) = \partial_k F_{-}(\Omega(k)) = 2k - \frac{2\alpha_{2}(\alpha_{1}\Omega(k) + \alpha_{2}k)}{\sqrt{(\alpha_{1}\Omega(k) + \alpha_{2}k)^{2} + \rho_{1}^{2}\rho_{2}^{2}}}.
\end{align*}
Combining with  $F_-'(\Omega(k)) = 0$, 
$$
\frac{\alpha_1\Omega(k) + \alpha_2 k}{\sqrt{(\alpha_1\Omega(k) + \alpha_2 k)^2 + \rho_1^2\rho_2^2}} = -\frac{\Omega(k)}{\alpha_1},
$$ 
this yields
\begin{align*}
m'(k) = 2k + \frac{2\alpha_2}{\alpha_1}\Omega(k) = 2k  \frac{\alpha_{1}^{2} - \alpha_{2}^{2} + \sqrt{(\alpha_{1}\Omega(k) + \alpha_{2}k)^{2} + \rho_{1}^{2}\rho_{2}^{2}}}{\alpha_{1}^{2} + \sqrt{(\alpha_{1}\Omega(k) + \alpha_{2}k)^{2} + \rho_{1}^{2}\rho_{2}^{2}}}.
\end{align*}
Since $\alpha_1^2 - \alpha_2^2 = 1$, we have $m'(k) > 0$ for every $k > 0$ 
so that the value of $k_0 \in (k_2,k_1)$, where $m(k_0) = 0$, is unique.

Let $\Omega_1 \in \mathbb{R}$ be the argument of $\max_{\Omega \in \mathbb{R}} F_{-}(\Omega)$ at $k = k_0$ such that $F_-(\Omega_1) = F_-'(\Omega_1) = 0$. By Lemma \ref{lem 3},
$F_+(\Omega_1) \neq 0$ so that $F_+(\Omega)$ admits only two real roots 
for every $|k| \geq k_0$, because the total number of roots of $F_+(\Omega)$ and $F_-(\Omega)$ cannot exceed four real roots of the quartic polynomial $P(\Omega)$. 

We now consider the real roots of $F_+(\Omega)$ for $|k| < k_0$, where $F_-(\Omega)$ does not have real roots. Since $F_+(\Omega_1) \neq 0$ at $k = k_0$ by Lemma \ref{lem 3} and $F_+(\Omega)$ has only two real roots at $k = k_0$, $F_+(\Omega)$ continues to have only two real roots for $k \lesssim k_0$. Hence, two roots of $P(\Omega)$ are complex-conjugate for $k \lesssim k_0$.
We proceed separately for $M > 0$ and $M < 0$.

\underline{Case $M > 0$.} By Lemma \ref{lem 1}, $F_+(\Omega) |_{k=0}$ admits a double zero root and no nonzero real roots. We show 
that $F_+(\Omega)$ admits only two simple real roots for any $k \neq 0$.
In view of the two real roots of $F_-(\Omega)$ for $|k| > k_0$, we 
obtain that $P(\Omega)$ has two real and two complex-conjugate roots for $|k| < k_0$ and four real roots for $|k| > k_0$.

To show that $F_+(\Omega)$ admits only two simple real roots for any $k \neq 0$, we prove that $F_+(\Omega)$ is a strictly convex function for $M > 0$. Differentiating $F_+$ in $\Omega$, we obtain
	\begin{align*}
F_+'(\Omega) = 2\Omega - \frac{2\alpha_1(\alpha_1\Omega + \alpha_2 k)}{\sqrt{(\alpha_1\Omega + \alpha_2 k)^2 + \rho_1^2\rho_2^2}}
	\end{align*}
and
	\begin{align*}
F_+''(\Omega) = 2 - \frac{2\alpha_1^2\rho_1^2\rho_2^2}{\left ( (\alpha_1\Omega + \alpha_2 k)^2 + \rho_1^2\rho_2^2 \right)^{3/2}}.
	\end{align*}
Since $(\alpha_1\Omega + \alpha_2 k)^2 + \rho_1^2\rho_2^2 \geq \rho_1^2\rho_2^2$ and 
\begin{align}
\label{eq-M}
M = 4\rho_1^3\rho_2^3 - (\rho_1^2+\rho_2^2)^2 = 4\rho_1^3\rho_2^3 - 4\alpha_1^2\rho_1^2\rho_2^2 = 4\rho_1^2\rho_2^2 \left(\rho_1\rho_2 - \alpha_1^2\right) > 0, 
\end{align}
we conclude that 
	\begin{align*}
F_+''(\Omega) \geq 2 - \frac{2\alpha_1^2 \rho_1^2 \rho_2^2}{\rho_1^3 \rho_2^3} = 2\left(1 - \frac{\alpha_1^2}{\rho_1\rho_2}\right) > 0.
	\end{align*}
We have previously proven that 	$F_+(\Omega_*) < 0$ for $k \neq 0$ and $F_+(\Omega) \to +\infty$ as $|\Omega| \to \infty$, from which $F_+(\Omega)$ has at least two real roots for any $k \neq 0$. Since $F_+(\Omega)$ is a strictly convex function for any $k \in \mathbb{R}$ if $M > 0$, it admits exactly two simple real roots for any $k \neq 0$. 
By Lemma \ref{lem 1}, the two real roots become as $k \to 0$ the double zero root of $F_+(\Omega) |_{k=0}$. Thus, $P(\Omega)$ has exactly four real roots for $|k| > k_0$ and two real 
and two complex conjugate roots for $|k| < k_0$. This yields the assertion for $M > 0$.
	
\underline{Case $M < 0$.}  By Lemma \ref{lem 1}, 
$F_+(\Omega) |_{k=0}$ has a double zero root and two simple real roots. 
By Lemma \ref{lem 2}, the double zero root of $F_+(\Omega) |_{k=0}$ splits for small $k \neq 0$ as a pair of complex-conjugate roots if $\rho_1 \rho_2 < 1$ and another pair of real roots if $\rho_1 \rho_2 > 1$. Therefore, $F_+(\Omega)$ admits only two real roots for small $k \neq 0$ if $\rho_1 \rho_2 < 1$ and four real roots for small $k \neq 0$ 
if $\rho_1 \rho_2 > 1$. Recall also that $F_+(\Omega)$ has exactly two 
real roots for $|k| \geq k_0$.

Since $F_+(-\Omega) = F_+(\Omega) |_{k \to -k}$, the number of real roots of $F_+(\Omega)$ depends only on $|k|$. We will show next that 
\begin{itemize}
\item If $\rho_1 \rho_2 > 1$, then there exists exactly one point $k_* \in (0,\infty)$, where $F_+(\Omega) |_{|k|=k_*}$ admits a double real root,
\item If $\rho_1 \rho_2 < 1$ and $\alpha^2_{2}<\frac{(1-\rho_{1}\rho_{2}) (1+\rho_{1}\rho_{2})^3}{(2\rho_{1}\rho_{2}-1)^3}$, then $F_+(\Omega)$ does not admit a double real root for every $k \neq 0$, 
\item If $\frac{1}{2} < \rho_1 \rho_2 < 1$ and $\alpha^2_{2}>\frac{(1-\rho_{1}\rho_{2}) (1+\rho_{1}\rho_{2})^3}{(2\rho_{1}\rho_{2}-1)^3}$, then there exist exactly two points $k_*^{(1)} > k_*^{(2)} > 0$, where $F_+(\Omega) |_{|k|=k_*^{(1,2)}}$ admits double real roots.
\end{itemize}
Since the number of real roots of $F_+(\Omega)$ can change only when a double real root of $F_+(\Omega)$ appears at some $k \neq 0$, this implies that 
\begin{itemize}
\item If $\rho_1 \rho_2 > 1$, then $P(\Omega)$ has exactly four real roots for $0 < |k| < k_*$ and $|k| > k_0$ and two real 
and two complex conjugate roots for $k_* < |k| < k_0$. This yields the assertion with $k_- := k_*$ and $k_+ := k_0$.

\item If $\rho_1 \rho_2 < 1$ and $\alpha^2_{2}<\frac{(1-\rho_{1}\rho_{2}) (1+\rho_{1}\rho_{2})^3}{(2\rho_{1}\rho_{2}-1)^3}$, then
$P(\Omega)$ has exactly two real roots and two complex conjugate roots 
for every $0 < |k| < k_0$ and four real roots for $|k| > k_0$. This yields the assertion with the same $k_0$.

\item If $\frac{1}{2} < \rho_1 \rho_2 < 1$ and $\alpha^2_{2}>\frac{(1-\rho_{1}\rho_{2}) (1+\rho_{1}\rho_{2})^3}{(2\rho_{1}\rho_{2}-1)^3}$, then $P(\Omega)$ 
has exactly two real roots and two complex-conjugate roots for 
$0 < |k| < k_*^{(2)}$ and $k_*^{(1)} < |k| < k_0$ and 
four real roots  for $k_*^{(2)} < |k| < k_*^{(1)}$ and $|k| > k_0$. 
This yields the assertion with $k_0 := k_*^{(2)}$, $k_- := k_*^{(1)}$, and $k_+ := k_0$.
\end{itemize} 

To study double real roots of $F_+(\Omega)$ for $k \neq 0$, we set 
\begin{align}
\label{eq-1}
F_+(\Omega) = \Omega^2 - k^2 + 2\rho_1\rho_2 - 2\sqrt{(\alpha_1\Omega + \alpha_2 k)^2 + \rho_1^2\rho_2^2} = 0
\end{align}
and
\begin{align}
\label{eq-2}
F_+'(\Omega) = 2\Omega - \frac{2\alpha_1(\alpha_1\Omega + \alpha_2 k)}{\sqrt{(\alpha_1\Omega + \alpha_2 k)^2 + \rho_1^2\rho_2^2}} = 0.
\end{align}
It follows from (\ref{eq-2}) that
\begin{align*}
\alpha_1\alpha_2 k = \Omega \left[ \sqrt{(\alpha_1\Omega + \alpha_2 k)^2 + \rho_1^2\rho_2^2} - \alpha_1^2 \right].
\end{align*}
If $\alpha_2 = 0$ and $\alpha_1 = 1$ (when $\rho_1 = \rho_2$), no double roots exist. Indeed, the choice $\Omega = 0$ from (\ref{eq-2}) implies 
$k = 0$ from (\ref{eq-1}), which is a contradiction with $k \neq 0$. The other choice  $\Omega^2 = 1 - \rho_1^2 \rho_2^2$ from (\ref{eq-2}) implies 
$-(1-\rho_1\rho_2)^2 - k^2 = 0$, which is also a contradiction. Hence, in what follows, we may assume that $\alpha_2 \neq 0$ (when $\rho_1 \neq \rho_2$). 

Squaring and using (\ref{eq-2}) again with $\alpha_1 > 0$ and $\alpha_2 \neq 0$ yields 
\begin{align}
\label{eq-3}
\Omega^2 = \frac{\alpha_1^2 (\alpha_1\Omega + \alpha_2 k)^2}{(\alpha_1\Omega + \alpha_2 k)^2 + \rho_1^2\rho_2^2}, \quad 
k^2 = \frac{(\alpha_1\Omega + \alpha_2 k)^2  \left[ \sqrt{(\alpha_1\Omega + \alpha_2 k)^2 + \rho_1^2\rho_2^2} - \alpha_1^2 \right]^2}{\alpha_2^2 [ (\alpha_1\Omega + \alpha_2 k)^2 + \rho_1^2\rho_2^2]}.
\end{align}
Substituting (\ref{eq-3}) into (\ref{eq-1}) and simplifying by taking a common denominator using $\alpha_2^2 = \alpha_1^2 - 1$, we obtain
\begin{align*}
0 =& -\frac{\sqrt{(\alpha_1\Omega + \alpha_2 k)^2 + \rho_1^2\rho_2^2} - \rho_1\rho_2}{\alpha_2^2 \left[(\alpha_1\Omega + \alpha_2 k)^2 + \rho_1^2\rho_2^2\right]}  \Bigg\{ \left[(\alpha_1\Omega + \alpha_2 k)^2 + \rho_1^2\rho_2^2\right]^{3/2} + (\rho_1\rho_2 - 2) \\
& \quad  \left[(\alpha_1\Omega + \alpha_2 k)^2 + \rho_1^2\rho_2^2\right] + \alpha_1^2 (1 - 2\rho_1\rho_2) \left[(\alpha_1\Omega + \alpha_2 k)^2 + \rho_1^2\rho_2^2 \right]^{1/2} + \alpha_1^2 \rho_1\rho_2 \Bigg\}.
\end{align*}
Since $\alpha_1\Omega + \alpha_2 k = 0$ in (\ref{eq-1}) and (\ref{eq-2}) yields $\Omega = k = 0$, which is excluded for $k_* \in (0,\infty)$, 
the first factor is non-zero and the condition for the existence of 
the double real root can be rewritten in the form:
\begin{align}
\label{pol-D}
0 = y^3 + (\rho_1\rho_2 - 2) y^2 + \alpha_1^2 (1 - 2\rho_1\rho_2) y + \alpha_1^2 \rho_1\rho_2 =: Q(y),
\end{align}
where $y:= \sqrt{(\alpha_1\Omega + \alpha_2 k)^2 + \rho_1^2\rho_2^2}$. 
Since $\sqrt{(\alpha_1\Omega + \alpha_2 k)^2 + \rho_1^2\rho_2^2} \geq \rho_1 \rho_2$ for $\Omega \in \mathbb{R}$ and $k \in \mathbb{R}$, we only need to check the number of roots of $Q(y)$ for $y \geq \rho_1 \rho_2$. 

\underline{Case $\rho_1\rho_2 > 1$.} We show that there exists exactly one point $k_* \in (0,\infty)$, where $F_+(\Omega) |_{|k|=k_*}$ admits a double real root. We compute from (\ref{pol-D}) that 
	\begin{align*}
Q(\rho_1\rho_2) = 2\rho_1\rho_2 (\rho_1\rho_2 - 1)(\rho_1\rho_2 - \alpha_1^2) < 0,
	\end{align*}
where the negative value holds because $\rho_1\rho_2 > 1$ and $\rho_1\rho_2 - \alpha_1^2 < 0$ for $M < 0$, see (\ref{eq-M}). We further obtain from (\ref{pol-D}) that 
	\begin{align*}
Q(\alpha_1^2) = \alpha_1^2 (\alpha_1^2 - 1)(\alpha_1^2 - \rho_1\rho_2) > 0,
	\end{align*}
where the positive value holds because $\alpha_1^2 > 1$ and $\alpha_1^2 > \rho_1\rho_2$. Therefore, $Q$ has at least one zero in $(\rho_1\rho_2,\alpha_1^2)$. Differentiating $Q(y)$ in $y$ yields
	\begin{align*}
Q'(y) = 3 y^2 + 2(\rho_1\rho_2 - 2) y + \alpha_1^2(1 - 2\rho_1\rho_2)
	\end{align*}
and
	\begin{align*}
Q''(y) = 6 y + 2(\rho_1\rho_2 - 2).
	\end{align*}
We have $Q''(y) \geq 8\rho_1\rho_2 - 4 > 0$ for $y \geq \rho_1\rho_2$ and $\rho_1 \rho_2 > 1$, so that $Q'(y)$ is strictly increasing for $y \geq \rho_1\rho_2$. Due to this strict increase of $Q'(y)$ with $Q(\rho_1\rho_2) < 0$ and $Q(\alpha_1^2) > 0$, there exists exactly one root of $Q$ in $(\rho_1\rho_2,\alpha_1^2)$ and no other roots of $Q(y)$ for $y \geq \alpha_1^2$. This root uniquely determines $\Omega^2$ and $k^2$ by (\ref{eq-3}) so that there exists exactly one point $k_* \in (0,\infty)$, where $F_+(\Omega) |_{|k|=k_*}$ admits a double real root and the assertion is proven. 

\underline{Case $\rho_1\rho_2 < 1$.} We continue to analyze real roots of $Q(y)$ given by (\ref{pol-D}) in the admissible region $y\geq \rho_1\rho_2$. Since $\rho_1\rho_2 < 1$ and $\rho_1 \rho_2 - \alpha_1^2 < 0$ for $M < 0$, we now have 
$$
Q(\rho_1 \rho_2) = 2 \rho_1 \rho_2 (\rho_1 \rho_2 - 1) (\rho_1 \rho_2 - \alpha_1^2) > 0,
$$
so that we have either no or two roots of $Q(y)$ for $y \geq \rho_1 \rho_2$.
Since $\alpha_1^2 = 1 + \alpha_2^2$, we can rewrite 
$Q(y)$ in the form:
\begin{align*}
Q(y)=(y-1)^2(y+\rho_1\rho_2)+\alpha_2^2(\rho_1\rho_2+(1-2\rho_1\rho_2)y),
\end{align*}
where $(y-1)^2 (y+\rho_1\rho_2) \geq 0$, $\alpha_2 \neq  0$, and $\rho_1 \rho_2+(1-2\rho_1\rho_2)y > 0$ for $y > \rho_1 \rho_2$ if $\rho_1 \rho_2 < \frac{1}{2}$. Therefore, no real roots of $Q(y)$ exist in the admissible region $y \geq \rho_1 \rho_2$ if $\rho_1 \rho_2 <\frac{1}{2}$. 

If $\frac{1}{2}<\rho_1\rho_2<1$, the existence of real roots of $Q(y)$ for $y \geq \rho_1 \rho_2$ implies that 
\begin{align*}
(2\rho_1\rho_2-1)y-\rho_1\rho_2>0, \quad \mbox{\rm that is,} \quad 
y>\frac{\rho_1\rho_2}{2\rho_1\rho_2-1}.
\end{align*}
The real root of $Q$ is equivalently found from
\begin{align}
\label{eq-4}
\alpha_2^2=\frac{(y-1)^2(y+\rho_1\rho_2)}{(2\rho_1\rho_2-1)y-\rho_1\rho_2}.
\end{align}
Differentiating the right-hand side of (\ref{eq-4}) with respect to $y$ yields
\begin{align*}
\frac{d}{dy}\left(\frac{(y-1)^2(y+\rho_1\rho_2)}{(2\rho_1\rho_2-1)y-\rho_1\rho_2}\right)=
\frac{2y(y-1)((2\rho_1\rho_2-1)y-\rho_1\rho_2(2-\rho_1\rho_2))}{(
	(2\rho_1\rho_2-1)y-\rho_1\rho_2)^2},
\end{align*}
with the only critical point at $y = y_*$ given by 
\begin{align*}
y_{*}=\frac{\rho_1\rho_2(2-\rho_1\rho_2)}{2\rho_1\rho_2-1},
\end{align*}
which satisfies the restrictions $y_* > 1$ and $y_* > \frac{\rho_1\rho_2}{2\rho_1\rho_2-1}$. Therefore, the right-hand side of (\ref{eq-4}) is strictly decreasing on $\left (\frac {\rho_1 \rho_2} {2 \rho_1 \rho_2-1}, y_* \right)$ and increasing on $(y_*, \infty) $ with the  unique minimum attained at
\begin{align*}
\frac{(1-\rho_1\rho_2)(1+\rho_1\rho_2)^3}{	(2\rho_1\rho_2-1)^3}.
\end{align*}
If $\alpha_2^2<\frac{(1-\rho_1\rho_2)(1+\rho_1\rho_2)^3}{(2\rho_1\rho_2-1)^3}$,
then no real roots of $Q(y)$ exist in the admissible region $y \geq \rho_1 \rho_2$ so that $F_+(\Omega)$ does not admit a double real root for every $k \neq 0$. If $\alpha_2^2 > \frac{(1-\rho_1\rho_2)(1+\rho_1\rho_2)^3}{(2\rho_1\rho_2-1)^3}$,
then two real roots of $Q(y)$ exist in the admissible region $y \geq \rho_1 \rho_2$ so that there exist exactly two points $k_*^{(1)} > k_*^{(2)} > 0$, where $F_+(\Omega) |_{k = k_*^{(1,2)}}$ admits double real roots. The assertion is proven in both cases. 
\end{proof}

\subsection{Examples of the nonzero background}

We finish this section with three examples of the nonzero, spatially homogeneous background (\ref{0-slo}) with $\ell = 0$.

\begin{figure}[htb!]
	\centering
	\includegraphics[width=0.7\columnwidth,height=6cm]{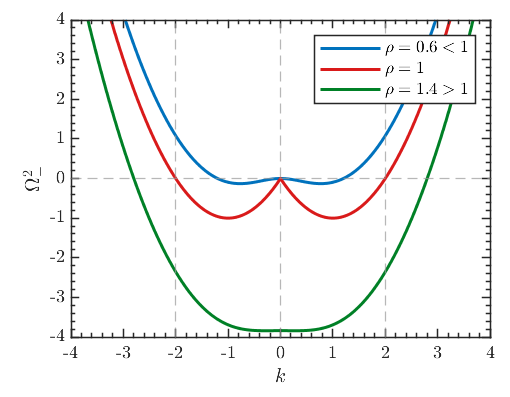}
	\caption{The values of $\Omega_-^{2}$ versus $k$ for $\rho_1 = \rho_2 \equiv \rho$ with $\rho=0.6$ (blue), $\rho=1$ (red), and $\rho=1.4$ (green). 
	}
	\label{fig-Omega}
\end{figure}

\begin{example}[$\rho_1=\rho_2 \equiv \rho$]

The characteristic polynomial $P(\Omega)$ is obtained from \eqref{characteristic3} as
\begin{equation*}
	P(\Omega) = \Omega^4 - \Omega^2(4+2 k^2-4 \rho^2)+k^2(k^2-4 \rho^2).
\end{equation*}
The bi-quadratic equation $P(\Omega) = 0$ yields two pairs of roots $\pm \Omega_+$, $\pm \Omega_-$, where $\Omega_{\pm}$ are given by 
\begin{equation*}
	\Omega^2_{\pm} = k^2 + 2(1 - \rho^2) \pm 2\sqrt{k^2 + (1 - \rho^2)^2}.
\end{equation*}
Since $\Omega_{+}^2 > 0$ for $k \neq 0$, we have $\Omega_+ \in \mathbb{R}$. The sign of $\Omega_-^2$ is found from 
$$
\Omega_{+}^2 \Omega_{-}^2 = k^2(k^2 - 4\rho^2),
$$
where $\Omega_+^2 > 0$ for $k \neq 0$. If $0 < |k| < 2 \rho$, then $\Omega_-^2 < 0$ so that $\Omega_- \in {\rm i} \mathbb{R}$ and $P(\Omega)$ has two real roots and two imaginary roots. If $|k| > 2 \rho$, then $\Omega_-^2 > 0$ so that $\Omega_- \in \mathbb{R}$ and $P(\Omega)$ has four real roots.
To distinguish between the modulational instability and the full-band instability in Theorem \ref{theorem-stab}, we compute $\Omega_{\pm}^2$ at $k = 0$:
$$
\Omega^2_{\pm} |_{k=0} = 2(1 - \rho^2) \pm 2|1 - \rho^2|.
$$ 
If $0 < \rho < 1$, then $\Omega_{+}^2 = 4(1 - \rho^2) > 0$ and $\Omega_{-}^2 = 0$. This is the modulational instability case in Theorem \ref{theorem-stab}. If $\rho > 1$, then $\Omega_{+}^2 = 0$ and $\Omega_{-}^2 = -4(\rho^2 - 1) < 0$. This is the full-band instability case in Theorem \ref{theorem-stab}. The graph of $\Omega_-^2$ versus $k$ for three different values of $\rho$ is shown in Figure \ref{fig-Omega}. The unstable bands correspond to $\Omega_-^2 < 0$. The difference between the modulational instability for $\rho < 1$ and the full-band instability for $\rho > 1$ is clearly observed from the plot. The marginal case $\rho= 1$ corresponds to the modulational instability.
\end{example}

\begin{figure}[htb!]
	\centering
	\includegraphics[width=0.7\columnwidth,height=7cm]{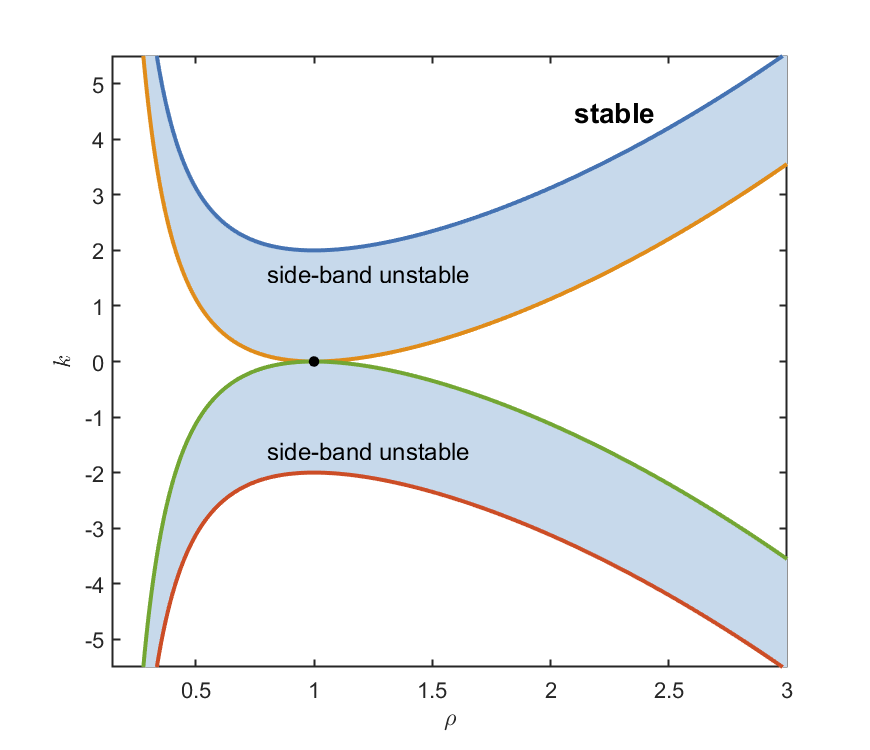}
	\caption{The two instability bands in the $(\rho,k)$ plane for $\rho_1 = \frac{1}{\rho_2} \equiv \rho$ are shown by the shaded regions.}
	\label{fig-rho}
\end{figure}

\begin{example}[$\rho_1=\frac{1}{\rho_2} \equiv \rho$]

The characteristic polynomial $P(\Omega)$ given by \eqref{characteristic3} 
can be factorized as
\begin{align*}	
	P(\Omega) = \left[\left(\Omega+\frac{1-\rho^4}{2 \rho^2}\right)^2 - \left(k+\frac{1+\rho^4}{2 \rho^2}\right)^2 + 1\right] \;  \left[\left(\Omega-\frac{1-\rho^4}{2 \rho^2}\right)^2 - \left(k-\frac{1+\rho^4}{2 \rho^2}\right)^2 + 1\right].
\end{align*}
Consequently, $P(\Omega) = 0$ admits four roots $\Omega_{1,\pm}$ and $\Omega_{2,\pm}$ given by roots of the two uncoupled quadratic equations:
\begin{align*}
	\Omega_{{1,\pm}} &= \frac{(\rho^4-1)}{2\rho^2}\pm \sqrt{\left(k+\frac{\rho^4+1}{2 \rho^2}\right)^2-1},\\
	\Omega_{{2,\pm}} &= -\frac{(\rho^4-1)}{2\rho^2}\pm \sqrt{\left(k-\frac{\rho^4+1}{2 \rho^2}\right)^2-1}.
\end{align*}
If 
$$
\left(k+\frac{1+\rho^4}{2\rho^2}\right)^2-1 < 0 \quad \Rightarrow \quad  \left|k+\frac{1+\rho^4}{2\rho^2}\right| < 1,
$$ 
then $\Omega_{1,\pm}$ are complex-conjugate. If 
$$
\left(k-\frac{1+\rho^4}{2\rho^2}\right)^2-1 < 0 \quad \Rightarrow \quad  \left|k-\frac{1+\rho^4}{2\rho^2}\right| < 1,
$$ 
then $\Omega_{2,\pm}$ are complex-conjugate. This yields two instability bands, 
one is located between $\left( -\frac{(1+\rho^2)^2}{2\rho^2},-\frac{(1-\rho^2)^2}{2\rho^2} \right)$ with the center at $k = -\frac{1+ \rho^4}{2 \rho^2}$ and the other one is located between $\left( \frac{(1-\rho^2)^2}{2\rho^2},\frac{(1+\rho^2)^2}{2\rho^2} \right)$ with the center at $k = \frac{1+ \rho^4}{2 \rho^2}$. Both instability bands are disjoint from each other except for $\rho = 1$, the boundaries of the two instability bands meet at $k = 0$ with $\Omega_{1,\pm} = \Omega_{2,\pm} = 0$ at $k = 0$. This is the side-band instability case in Theorem \ref{theorem-stab}. 

Figure \ref{fig-rho} displays the two instability bands in the $(\rho,k)$ plane. The case $\rho_1 \rho_2 = 1$ is marginal in Lemma \ref{lem 2} since the asymptotic behavior of $\Omega_{\pm}(k)$ for small $k$ in (\ref{def-beta}) admits $\beta_+ = \beta_-$ from the two roots of the 
quadratic equation (\ref{quad-beta}). The marginal case breaks as the modulational instability if $\rho_1\rho_2 < 1$ and as the side-band instability if $\rho_1 \rho_2 > 1$. The marginal case $\rho_1 \rho_2 = 1$ belongs to the side-band instability with the exception of $\rho_1 = \rho_2 = 1$, for which the two side bands connect at $k = 0$ and the marginal case belongs to the modulational instability.
\end{example}

\begin{example}[$\rho_1=-\rho_2 \equiv \rho$]
	
The characteristic polynomial $P(\Omega)$ is obtained from \eqref{characteristic3} as
\begin{equation*}	
	P(\Omega) = \Omega^4 - \Omega^2(4 + 2k^2 + 4 \rho^2) +k^2 (k^2 + 4\rho^2).
\end{equation*}
The bi-quadratic equation $P(\Omega) = 0$ yields two pairs of roots $\pm \Omega_+$, $\pm \Omega_-$, where $\Omega_{\pm}$ are given by 
$$
\Omega^2_{\pm} = k^2 +2(1+\rho^2) \pm 2\sqrt{k^2 + (1 + \rho^2)^2}.
$$
Since $k^2 + 2(1+\rho^2) \geq 2\sqrt{k^2 + (1+\rho^2)^2}$, we have 
$\Omega_+^2 \geq 0$ and $\Omega_-^2 \geq 0$ so that $P(\Omega)$ admits four real 
roots and the background is stable, in agreement with Theorem \ref{theorem-stab}.
\end{example}

\section{Algebraic dark solitons and rogue waves}
\label{sec-dark}

\subsection{Bilinear equations for solutions on the nonzero background} 

To prove Theorem \ref{theorem-soliton}, we first rewrite the MTM system (\ref{MTM}) on the nonzero background (\ref{0-slo})	by using bilinear equations, as in \cite{Chen-SAPM-2023,Baofeng1}. We represent the solutions in the form:
\begin{equation}
	\label{transfor}
	\begin{cases}
	u(x,t) = \rho_1 e^{{\rm i} (\ell x-\omega t)} \tilde{u}(x,t), \\
	v(x,t) = \rho_2 e^{{\rm i} (\ell x-\omega t)} \tilde{v}(x,t),
	\end{cases}
\end{equation}
where $\ell$, $\omega$ are defined by (\ref{relation}) and  $\rho_1$, j$\rho_2$ are real parameters. Substituting (\ref{transfor}) into \eqref{MTM} and using \eqref{relation} to eliminate $\ell$ and $\omega$, we obtain
\begin{align}\label{system}
	\begin{cases}
		&	{\rm i}(\tilde{u}_t+\tilde{u}_x)+\frac{\rho_2}{\rho_1}(\tilde{v}-\tilde{u})
		=
		\rho_2^2\,\tilde{u}\bigl(|\tilde{v}|^2-1\bigr),\\
		&	{\rm i}(\tilde{v}_t-\tilde{v}_x)+\frac{\rho_1}{\rho_2}(\tilde{u}-\tilde{v})
		=
		\rho_1^2\,\tilde{v}\bigl(|\tilde{u}|^2-1\bigr).
	\end{cases}
\end{align}
We are looking for solutions of (\ref{system}) satisfying $|\tilde{u}(x,t)|, |\tilde{v}(x,t)| \to 1$ as $|x| \to \infty$ for any $t \in \mathbb{R}$. 
Using   
\begin{equation}
\label{Hirota}
	\tilde{u}=\frac{g}{\bar{f}},\quad \tilde{v}=\frac{h}{f},
\end{equation}
with complex-valued functions $f(x,t)$, $g(x,t)$, and $h(x,t)$, we rewrite \eqref{system} in the form
\begin{equation}
\label{Hirota-trilinear}
	\begin{cases}
		& \frac{{\rm i} (D_t+D_x)(g\cdot \bar f)}{\bar f^{\,2}}
		+\frac{\rho_2}{\rho_1}\left(\frac{h}{f}-\frac{g}{\bar f}\right)
		=
		\rho_2^2\frac{g}{\bar f}
		\left(\frac{h\bar h}{f\bar f}-1\right),\\
		& \frac{{\rm i} (D_t-D_x)(h\cdot f)}{f^{2}}
		+\frac{\rho_1}{\rho_2}\left(\frac{g}{\bar f}-\frac{h}{f}\right)
		=
		\rho_1^2\frac{h}{f}
		\left(\frac{g\bar g}{f\bar f}-1\right),	
		\end{cases}
\end{equation}
where $D_t$ and $D_x$ are standard Hirota operators. Following \cite{Chen-SAPM-2023,Baofeng1}, the under-determined system  (\ref{Hirota-trilinear}) can be rewritten as the following over-determined system of four bilinear equations:
\begin{equation}
\label{Hirota-bilinear}
\begin{cases}
{\rm i} (D_t + D_x)(g \cdot f)+\frac{\rho_{2}}{\rho_{1}} (h \bar{f}-gf)=0,\\
{\rm i} (D_t - D_x) (h \cdot \bar{f})+\frac{\rho_{1}}{\rho_{2}}(g f- h \bar{f})=0,\\
{\rm i} (D_t + D_x)(f \cdot \bar{f})+\rho^2_{2} (|f|^2-|h|^2)=0,\\
{\rm i} (D_t - D_x)(\bar{f}\cdot f)+\rho^2_{1}(|f|^2-|g|^2)=0.
\end{cases}
\end{equation}
The proof of Theorem \ref{theorem-soliton} is obtained by the explicit 
construction of the algebraic dark solitons and rogue waves in 
the system (\ref{Hirota-bilinear}), based on Propositions \ref{prop-algebraic}, \ref{prop-velocity}, and \ref{prop-rogue}.

\subsection{Algebraic dark solitons}

The following proposition gives the explicit expressions for the traveling wave solutions (\ref{TW-dark}) with the rational functions $\mathfrak{u}_{\pm}$, $\mathfrak{v}_{\pm}$ satisfying $|\mathfrak{u}_{\pm}-1|, |\mathfrak{v}_{\pm}-1| = \mathcal{O}(|x|^{-1})$ as $|x| \to \infty$. 

\begin{prop}
	\label{prop-algebraic}
	If either $\rho_1 \rho_2 < 0$ or $\rho_1 \rho_2 > 1$ and $(\rho_1^2 + \rho_2^2)^2 \neq 4 \rho_1^3 \rho_2^3$, there exist exactly two solutions of the bilinear equations (\ref{Hirota-bilinear}) in the form:
\begin{align}
\label{dark soliton}
\begin{cases}
f = x - c t + {\rm i} \alpha_f, \\
g = x - c t + {\rm i} \alpha_g, \\ 
h = x - c t + {\rm i} \alpha_h,
\end{cases}
\end{align}	
where  $c = c_{\pm}$ takes one of the two admissible values given by 
\begin{equation}
\label{speed-c}
c_{\pm} = \frac{\rho_{1}^{4} - \rho_{2}^{4} \pm 4 \rho_{1}^{2} \rho_{2}^{2} \sqrt{\rho_1 \rho_2 (\rho_1 \rho_2 - 1)}}{(\rho_{1}^{2}+\rho_{2}^{2})^{2} - 4 \rho_{1}^{3} \rho_{2}^{3}}
\end{equation}
and $\alpha_f,\alpha_g,\alpha_h \in \mathbb{R}$ are uniquely defined 
for each value of $c  = c_{\pm}$. The two solutions of the MTM system (\ref{MTM}) are uniquely defined in the form (\ref{TW-dark}) by (\ref{transfor}), (\ref{Hirota}), and (\ref{dark soliton}) modulo the translational and rotational symmetry transformations.
\end{prop}

\begin{proof}
Substituting \eqref{dark soliton} into \eqref{Hirota-bilinear} yields the following system of equations:
\begin{align}
\label{algebraic-eqs1}
\begin{cases}
(1-c)(\alpha_g - \alpha_f) + \frac{\rho_2}{\rho_1} [ {\rm i} (x-ct) (\alpha_h - \alpha_g - 2 \alpha_f) + (\alpha_h + \alpha_g) \alpha_f] = 0, \\
-(1+c)(\alpha_h + \alpha_f) + \frac{\rho_1}{\rho_2} [ {\rm i} (x-ct) (\alpha_g - \alpha_h + 2 \alpha_f) - (\alpha_h + \alpha_g) \alpha_f] = 0, \\
2(1-c) \alpha_f + \rho_2^2(\alpha_f^2 - \alpha_h^2) = 0, \\
2(1+c) \alpha_f + \rho_1^2(\alpha_f^2 - \alpha_g^2) = 0. \\
\end{cases}
\end{align}
The linear term in the system (\ref{algebraic-eqs1}) in $x-ct$ is identically equal to $0$ if $\alpha_h = \alpha_g + 2 \alpha_f$. We require  $\alpha_f \neq 0$ for a bounded, nontrivial solution. Substituting $\alpha_h = \alpha_g + 2 \alpha_f$ into the constant terms of the first two equations in \eqref{algebraic-eqs1}, we obtain
\begin{align}
\label{alg-term}
\begin{cases}
&(1-c)(\alpha_g-\alpha_f)+2\frac{\rho_2}{\rho_1}\alpha_f(\alpha_f+\alpha_g)=0,\\
&(1+c)(\alpha_g+3\alpha_f)+2\frac{\rho_1}{\rho_2}\alpha_f(\alpha_f+\alpha_g)=0.
\end{cases}
\end{align}
The requirement $\alpha_f \neq 0$ implies $\alpha_f + \alpha_g \neq 0$. Indeed, if $\alpha_f \neq 0$ and $\alpha_f + \alpha_g = 0$, then 
the system (\ref{alg-term}) yields a contradiction with two different values $c = 1$ and $c = -1$ for the same parameter $c$. Hence, $\alpha_f \neq 0$ and $\alpha_f + \alpha_g \neq 0$. 

A rearrangement of the system (\ref{alg-term}) with  $\alpha_{g}-\alpha_{f}=(\alpha_{f}+\alpha_{g})-2\alpha_{f}$ and $\alpha_{g}+3\alpha_{f}=(\alpha_{f}+\alpha_{g})+2\alpha_{f}$ yields
\begin{align}\label{algebraic-eqs2}
\begin{cases}
(1-c)(\alpha_{f}+\alpha_{g})+2 \alpha_{f} \left[\frac{\rho_{2}}{\rho_{1}}(\alpha_{f}+\alpha_{g})-(1-c)\right]=0, \\
(1+c)(\alpha_{f}+\alpha_{g})+2 \alpha_{f} \left[\frac{\rho_{1}}{\rho_{2}}(\alpha_{f}+\alpha_{g})+(1+c)\right]=0.
\end{cases}
\end{align}
Multiplying the first equation of the system (\ref{algebraic-eqs2}) by $\frac{\rho_1}{\rho_2} (\alpha_f + \alpha_g) + (1+c)$ and the second equation of the system by $\frac{\rho_2}{\rho_1} (\alpha_f + \alpha_g) - (1-c)$, subtracting one from another, and dividing by $\alpha_f + \alpha_g \neq 0$, we obtain
\begin{align}\label{discriminant-eq1}
\left[(1-c)\frac{\rho_{1}}{\rho_{2}}-(1+c)\frac{\rho_{2}}{\rho_{1}}\right](\alpha_{f}+\alpha_{g}) = 2(c^2-1).
\end{align}

Now, we turn to the last two equations of the system (\ref{algebraic-eqs1}).  Expressing $(\alpha_g - \alpha_f)$ from the 
first equation of the system (\ref{alg-term}) and substituting it 
to the fourth equation of the system (\ref{algebraic-eqs1}) yields 
\begin{align}\label{algebraic-eqs3}
(\alpha_{f}+\alpha_{g})^2=\frac{c^2-1}{\rho_{1}\rho_{2}}.
\end{align}
Expressing $\alpha_h + \alpha_f = \alpha_g + 3 \alpha_f$ from the 
second equation of the system (\ref{alg-term}) and substituting it 
to the third equation of the system (\ref{algebraic-eqs1}) yields 
the same equation (\ref{algebraic-eqs3}), hence, it is a redundant equation. Substituting $(c^2-1)$ from (\ref{algebraic-eqs3}) into the right-hand side of (\ref{discriminant-eq1}) and dividing by 
$\alpha_f + \alpha_g \neq 0$ yields 
\begin{align}
\label{algebraic-eqs4}
\alpha_f+\alpha_g=\frac{(1-c)\rho_1^2  -  (1+c) \rho_2^2}{2\rho_1^2\rho_2^2}.
\end{align}
Substituting (\ref{algebraic-eqs4})  into \eqref{discriminant-eq1} 
yields a quadratic equation for $c$:
\begin{align}\label{velocity-eq}
\left[ (1-c)\rho_1^2  -  (1+c) \rho_2^2 \right]^2-4\rho^3_{1}\rho^3_{2}(c^2-1)=0,
\end{align}
The coefficient in front of $c^2$ is nonzero if $(\rho_1^2 + \rho_2^2)^2 \neq 4 \rho_1^3 \rho_2^3$ and two roots exist. The discriminant of the quadratic equation is simplified to the form:
\begin{align*}
\Delta &= 4 (\rho_1^4 - \rho_2^4)^2 - 4 [(\rho_1^2 - \rho_2^2)^2 + 4 \rho_1^3 \rho_2^3] [(\rho_1^2 + \rho_2^2)^2 - 4 \rho_1^3 \rho_2^3]\\
&= 64 \rho_1^5 \rho_2^5 (\rho_1\rho_2-1).
\end{align*}
Since $\Delta > 0$ if either $\rho_1\rho_2<0$ or $\rho_1\rho_2>1$, there exist two real distinct roots for $c = c_{\pm}$, where $c_{\pm}$ is given by (\ref{speed-c}).

It remains to find the values of $\alpha_f$, $\alpha_g$, and $\alpha_h$ uniquely in terms of $c = c_{\pm}$. By expressing $(\alpha_f + \alpha_g)$ from \eqref{algebraic-eqs4} and substituting the result into the first equation of the system \eqref{algebraic-eqs2} yields the explicit expression for $\alpha_f$:
\begin{align}\label{alpha-f1}
	\alpha_{f}=-\frac{(1-c) \rho_{1} [(1-c) \rho^2_{1}-(1+c) \rho^2_{2}]}{2 \rho_{2} [(1-c) \rho^2_{1}-(1+c)\rho^2_{2}-2(1-c) \rho^3_{1} \rho_{2}]}.
\end{align}
The quadratic equation \eqref{velocity-eq} can be rewritten in the equivalent form:
\begin{align*}
0 &= \left[(1-c) \rho^2_{1} - (1+c) \rho^2_{2} \right]^2 + 2\rho^3_{1}\rho_{2}(1-c) \left[2 \rho^2_{2}(1+c)\right]\\
&= \left[(1-c) \rho^2_{1} - (1+c) \rho^2_{2} \right]^2 + 2\rho^3_{1}\rho_{2}(1-c)\left[ (1-c)  \rho^2_{1}+(1+c)\rho^2_{2} - ((1-c) \rho^2_{1} - (1+c) \rho^2_{2})\right].
\end{align*}
Dividing both sides by $(1-c) \rho^2_{1} - (1+c) \rho^2_{2} \neq 0$, we obtain
\begin{align}\label{relation1}
(1-c) \rho^2_{1}-(1+c)\rho^2_{2}-2(1-c) \rho^3_{1} \rho_{2}=-\frac{2 \rho^3_{1} \rho_{2} (1-c) \left[(1-c) \rho^2_{1}+(1+c) \rho^2_{2}\right]}{(1-c) \rho^2_{1} - (1+c) \rho^2_{2}}.
\end{align}
Substituting \eqref{relation1} into \eqref{alpha-f1} and using \eqref{velocity-eq} yields
\begin{align}
\alpha_{f} = \frac{[(1-c) \rho^2_{1}-(1+c) \rho^2_{2}]^2}{4 \rho^2_{1}\rho^2_{2} \left[(1-c) \rho^2_{1}+(1+c) \rho^2_{2}\right]} = \frac{\rho_{1}\rho_{2}(c^2-1)}{(1-c)\rho^2_{1}+(1+c)\rho^2_{2}}.
		\label{alpha-f2}
\end{align}
The expression for $\alpha_g$ is obtained from (\ref{algebraic-eqs4}) and 
the expression for $\alpha_h$ is obtained from $\alpha_h = \alpha_g + 2 \alpha_f$:
\begin{align}
\label{alpha-g}
\alpha_{g}&= \frac{(1-c) \rho^2_{1} - (1+c) \rho^2_{2}}{2 \rho_1^2 \rho_2^2} - \alpha_{f} = \frac{(1-c)^2 \rho^4_{1} -(1+c)^2\rho^4_{2}-2 \rho^3_{1}\rho^3_{2}(c^2-1)}{2 \rho^2_{1} \rho^2_{2} [(1-c) \rho^2_{1}+(1+c)\rho^2_{2}]}
\end{align}
and
\begin{align}
\label{alpha-h}
\alpha_{h}&= \frac{(1-c) \rho^2_{1} - (1+c) \rho^2_{2}}{2 \rho_1^2 \rho_2^2} + \alpha_{f} = \frac{(1-c)^2 \rho^4_{1} -(1+c)^2\rho^4_{2}+2 \rho^3_{1}\rho^3_{2}(c^2-1)}{2 \rho^2_{1} \rho^2_{2} [(1-c) \rho^2_{1}+(1+c)\rho^2_{2}]}.
\end{align}
These explicit expressions conclude the proof of the assertion.
\end{proof}

\begin{remark}
	The two values of the wave speed $c = c_{\pm}$ in (\ref{speed-c}) recover (\ref{speeds}).	In the exceptional case $(\rho_1^2 + \rho_2^2)^2 = 4 \rho_1^3 \rho_2^3$ for $\rho_1 \rho_2 > 1$, the coefficient in front of $c^2$ in the quadratic equation (\ref{velocity-eq}) vanishes and there exists only one admissible speed of the linear equation given by 
$$
c=\frac{(\rho^2_{1}-\rho^2_{2})^2+4\rho^3_{1} \rho^3_{2}}{2 (\rho^2_{1}+\rho^2_{2})  (\rho^2_{1}-\rho^2_{2}) },
$$
where $\rho_2 \neq \rho_1$ if $\rho_1 \rho_2 > 1$ and $(\rho_1^2 + \rho_2^2)^2 = 4 \rho_1^3 \rho_2^3$. The expressions for $\alpha_f$, $\alpha_g$, and $\alpha_h$ remain the same but with the uniquely defined value of  $c$, which recovers (\ref{speed-unique}).
\end{remark}

\begin{remark}
To characterize the spatial profiles of the algebraic dark solitons, we compute the value of their normalized intensity 
from (\ref{Hirota}) and (\ref{dark soliton}):
\begin{align*}
|\mathfrak{u}_{\pm}(x-ct)|^2 + |\mathfrak{v}_{\pm}(x-ct)|^2 &= 
\frac{2 (x-ct)^2 + \alpha_g^2 + \alpha_h^2}{(x-ct)^2 + \alpha_f^2} \\
&= 2 + \frac{[(1-c) \rho^2_{1} - (1+c) \rho^2_{2}]^2}{2 \rho_1^4 \rho_2^4 [(x-ct)^2 + \alpha_f^2]} \\
&= 2 + \frac{2 (c^2-1)}{\rho_1 \rho_2 [(x-ct)^2 + \alpha_f^2]},
\end{align*}
where we have used (\ref{alpha-g}) and (\ref{alpha-h}) as well as the quadratic equation (\ref{velocity-eq}). The normalized total intensity for their spatial profiles is of the elevation type with the maximum attained at the point of symmetry:
\begin{align*}
|\mathfrak{u}_{\pm}(0)|^2 + |\mathfrak{v}_{\pm}(0)|^2 &= 2 + \frac{2(c^2-1)}{\rho_1 \rho_2 \alpha_f^2} \\
&= 2 + 2 \frac{[(1-c) \rho_1^2 + (1+c) \rho_2^2]^2}{\rho_1^3 \rho_2^3 (c^2 - 1)} \\
&= 2 + 2 \frac{[(1-c) \rho_1^2 - (1+c) \rho_2^2]^2 - 4(c^2-1) \rho_1^2 \rho_2^2}{\rho_1^3 \rho_2^3 (c^2 - 1)} \\
&= 2 + 8 \frac{\rho_1 \rho_2 - 1}{\rho_1 \rho_2},
\end{align*}
where we have used (\ref{alpha-f2}) and again (\ref{velocity-eq}). 
This expression recovers (\ref{max-elevation-dark}). It is remarkable that the maximal intensity is independent of the wave speed $c$ and hence, it is identical for both algebraic dark solitons in Proposition \ref{prop-algebraic}.
\end{remark}

Next, we prove that the two values (\ref{speed-c}) in Proposition \ref{prop-algebraic} correspond to the end points of the admissible range of the wave speed $c$ of the exponentially decaying dark solitons. We consider traveling wave solutions of the system (\ref{system}) in the form:
\begin{equation}\
	\label{traveling wave}
\tilde{u}(x,t)=\mathfrak{u}(\xi),\quad \tilde{v}(x,t)=\mathfrak{v}(\xi), \quad 	\xi=x-c t,
	\end{equation}
subject to the boundary conditions: $|\mathfrak{u}(\xi)|, |\mathfrak{v}(\xi)| \to 1$ as $|\xi| \to \infty$. Substituting \eqref{traveling wave} into \eqref{system} yields the system of two first-order equations for complex-valued amplitudes:
	\begin{align}
	\label{ode-system}
	\begin{cases}
	{\rm i} (1-c) \mathfrak{u}^{\prime}+\frac{\rho_{2}}{\rho_{1}}(\mathfrak{v}-\mathfrak{u})=\rho^2_{2} \mathfrak{u}(|\mathfrak{v}|^2-1),\\
	-{\rm i} (1+c) \mathfrak{v}^{\prime}+\frac{\rho_{1}}{\rho_{2}}(\mathfrak{u}-\mathfrak{v})=\rho^2_{1}\mathfrak{v}(|\mathfrak{u}|^2-1).
	\end{cases}
	\end{align}
Due to the integrability, the system (\ref{ode-system}) can be solved explicitly by using the exponential functions \cite{Chen-SAPM-2023}. However, the explicit solutions are not the subject of our analysis here. Instead of solving the system (\ref{ode-system}), we only consider the admissible range of the wave speed $c$ for fixed $\rho_1,\rho_2 \in \mathbb{R}$. We show that the end points of the admissible range coincide with the values $c_+$ and $c_-$ defined by (\ref{speed-c}).
The corresponding result is given by the following proposition. 
	
	\begin{prop}
		\label{prop-velocity}
The exponentially decaying solutions of the system (\ref{ode-system}) with the boundary conditions $|\mathfrak{u}(\xi)|, |\mathfrak{v}(\xi)| \to 1$ as $|\xi| \to \infty$ exist only if 
\begin{itemize}
	\item $c \in (c_-,c_+)$ for either  $\rho_1 \rho_2 < 0$ or $\rho_1 \rho_2 > 1$ and $4 \rho_1^3 \rho_2^3 < (\rho_1^2 + \rho_2^2)^2$,
\end{itemize}
or
\begin{itemize}
	\item $c \in (-\infty,c_+) \cup (c_-,\infty)$ for $4 \rho_1^3 \rho_2^3 > (\rho_1^2 + \rho_2^2)^2$,
\end{itemize}
where $c_{\pm}$ are given by (\ref{speed-c}).
	\end{prop}

\begin{proof}
	The system (\ref{ode-system}) can be reduced to the planar Hamiltonian system by using the polar form $\mathfrak{u} = z_u e^{{\rm i} \theta_u}$, $\mathfrak{v} = z_v e^{{\rm i} \theta_v}$,
	with real-valued $z_u, z_v > 0$ and $\theta_u, \theta_v \in \mathbb{R} / 2\pi \mathbb{Z}$. The system (\ref{ode-system}) is rewritten in the form:
	\begin{align}
	\label{ode-system-polar}
\begin{cases}
(1-c) z_u' + \frac{\rho_2}{\rho_1} z_v \sin(\theta_v - \theta_u) = 0, \\
-(1-c) \theta_u' + \frac{\rho_{2} z_v}{\rho_{1} z_u} \cos(\theta_v - \theta_u) - \frac{\rho_{2}}{\rho_{1}}=\rho^2_{2} (z_v^2-1),\\
-(1+c) z_v' + \frac{\rho_{1}}{\rho_{2}} z_u \sin(\theta_u - \theta_v) = 0, \\
(1+c) \theta_v' + \frac{\rho_{1} z_u}{\rho_{2} z_v} \cos(\theta_u - \theta_v)  - \frac{\rho_{1}}{\rho_{2}} = \rho^2_{1} (z_u^2-1).
\end{cases}
\end{align}
Multiplying the first equation by $2 \rho_1^2 z_u$ and adding the third equation multiplied by $2 \rho_2^2 z_v$ shows that there exists a conserved quantity:
\begin{equation}
\label{conserved-ode}
(1-c) \rho_1^2 z_u^2 - (1+c) \rho_2^2 z_v^2 = (1-c) \rho_1^2 - (1+c) \rho_2^2,
\end{equation}
where we have used the boundary conditions: $z_u(\xi),z_v(\xi) \to 1$ as $|\xi| \to \infty$. In addition, the vector field of the system (\ref{ode-system-polar}) depends only on the phase difference $\theta := \theta_v - \theta_u$. Hence, the system (\ref{ode-system-polar}) can be closed as the second-order Hamiltonian system for $z_u$ and $\theta$, whereas $z_v$ is expressed from (\ref{conserved-ode}) and $\theta_u + \theta_v$ is found from integrating an explicit function of $\xi$ computed from the solution $z_u(\xi)$ and $\theta(\xi)$.
		
Equilibrium points of the system (\ref{ode-system-polar}) include 
the line $(z_u,z_v,\theta_u,\theta_v) = (1,1,\theta_0,\theta_0)$ with arbitrary $\theta_0 \in \mathbb{R}$. The exponentially decaying solutions to the system 
(\ref{ode-system}) with the boundary conditions $|\mathfrak{u}(\xi)|, |\mathfrak{v}(\xi)| \to 1$ as $|\xi| \to \infty$ exist only if the equilibrium point $(z_u,z_v,\theta_u,\theta_v) = (1,1,\theta_0,\theta_0)$ is a saddle point of the system (\ref{ode-system-polar}). Linearizing the system (\ref{ode-system-polar}) at $(1,1,\theta_0,\theta_0)$ and using the linear modes, e.g. $z_u(\xi) = \hat{z}_u e^{\kappa \xi}$, we reduce the linearized system to the characteristic equation given by 
		\begin{align*}
0 &= \begin{vmatrix}
(1-c) \kappa & 0 & -\frac{\rho_{2}}{\rho_{1}} & \frac{\rho_{2}}{\rho_{1}}\\
0 & -(1+c) \kappa & \frac{\rho_1}{\rho_2} & -\frac{\rho_1}{\rho_2} \\
-\frac{\rho_{2}}{\rho_{1}} & \frac{\rho_{2}}{\rho_{1}} - 2 \rho_2^2 & - (1-c) \kappa & 0 \\
\frac{\rho_1}{\rho_2} - 2 \rho_1^2 & -\frac{\rho_1}{\rho_2} & 0 & (1+c) \kappa 
		\end{vmatrix}\\
		&= \kappa^2  \left( \kappa^2 (1-c^2)^2   + \frac{\rho_1^2}{\rho_2^2} (1-c)^2 + \frac{\rho_2^2}{\rho_1^2} (1+c)^2 - 2 (1- 2 \rho_1 \rho_2) (1-c^2) \right).
		\end{align*}
The double zero eigenvalue of the characteristic equation is associated 
with the conserved quantity (\ref{conserved-ode}) and the uncoupled equation for $\theta_u + \theta_v$. The two nonzero roots determine the type of the equilibrium point $(z_u,z_v,\theta_u,\theta_v) = (1,1,\theta_0,\theta_0)$. The equilibrium point is a saddle point under the following condition:
$$
\frac{\rho_1^2}{\rho_2^2} (1-c)^2 + \frac{\rho_2^2}{\rho_1^2} (1+c)^2 - 2 (1- 2 \rho_1 \rho_2) (1-c^2) < 0.
$$
Multiplying by $\rho_1^2 \rho_2^2$, we recover 
$$
((1-c) \rho_1^2 - (1+c) \rho_2^2)^2 - 4 \rho_1^3 \rho_2^3 (c^2 - 1) < 0, 
$$
where the left-hand side is the same as in the quadratic equation (\ref{velocity-eq}). Since roots of the quadratic equation are given by $c_{\pm}$ in (\ref{speed-c}) 
and the $c^2$ term has the coefficient $(\rho_1^2 + \rho_2^2)^2 - 4 \rho_1^3 \rho_2^3$, we obtain the assertion from 
the fact that the quadratic curve is convex for either $\rho_1 \rho_2 < 0$ or $\rho_1 \rho_2 > 1$ and $4 \rho_1^3 \rho_2^3 < (\rho_1^2 + \rho_2^2)^2$ 
with $c_- < c_+$ and it is concave for $4 \rho_1^3 \rho_2^3 > (\rho_1^2 + \rho_2^2)^2$ with $c_+ < c_-$ .
\end{proof}

\begin{figure}[htb!]
	\centering
	\begin{minipage}[t]{0.45\columnwidth}
		\centering
		\includegraphics[width=\textwidth,height=6cm]{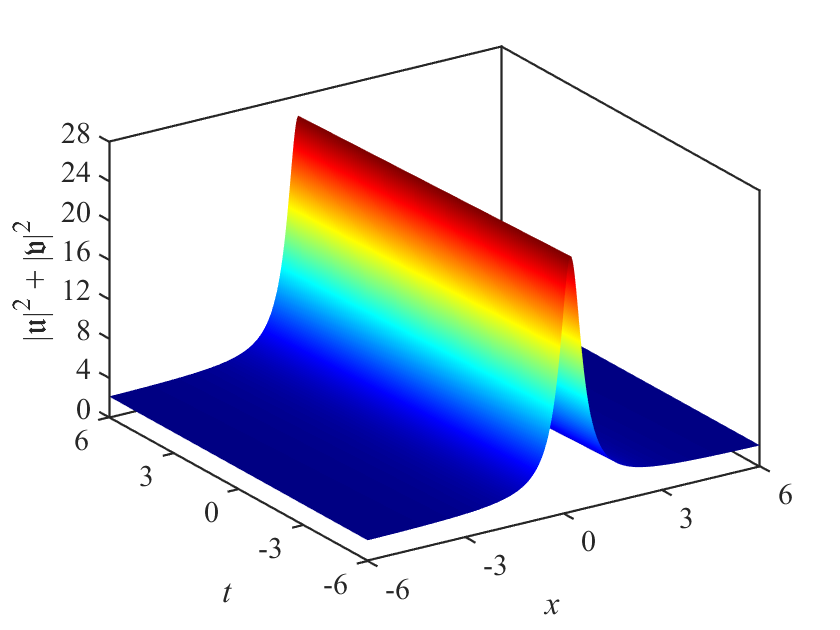}
		\vspace{0.1cm} 
		(a) 
	\end{minipage}
	\hfill
	\begin{minipage}[t]{0.45\columnwidth}
		\centering
		\includegraphics[width=\textwidth,height=6cm]{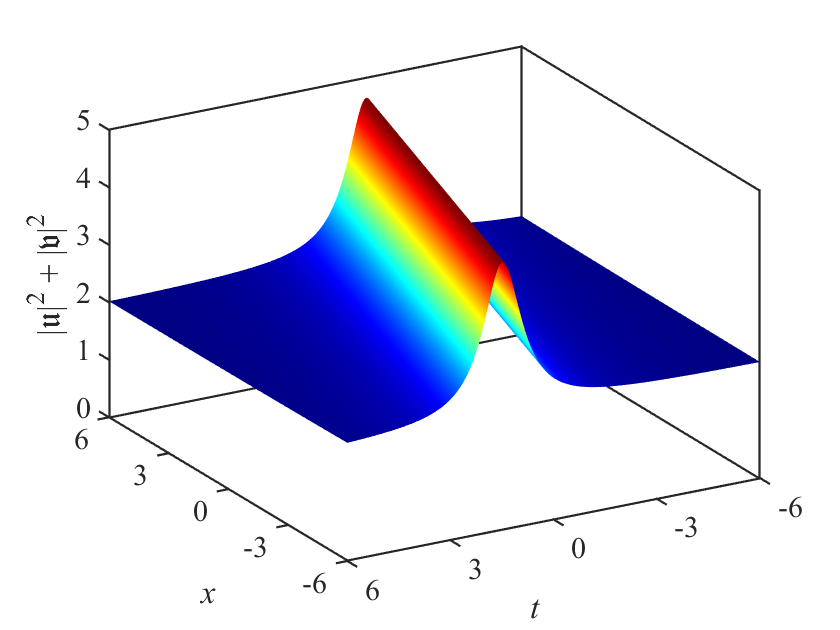}
		\vspace{0.1cm}
		(b) 
	\end{minipage}
	\caption{The total intensity $|\mathfrak{u}|^2 + |\mathfrak{v}|^2$ versus $(x,t)$ of the algebraic dark soliton with $c = c_-$ for (a) $\rho_{1}=0.5$, $\rho_{2}=-1$ and (b) $\rho_1 = 1.5$, $\rho_2 = 1$.}
	\label{FIG:9}
\end{figure}

\begin{figure}[htb!]
	\centering
	\begin{minipage}[t]{0.45\columnwidth}
		\centering
		\includegraphics[width=\textwidth,height=6cm]{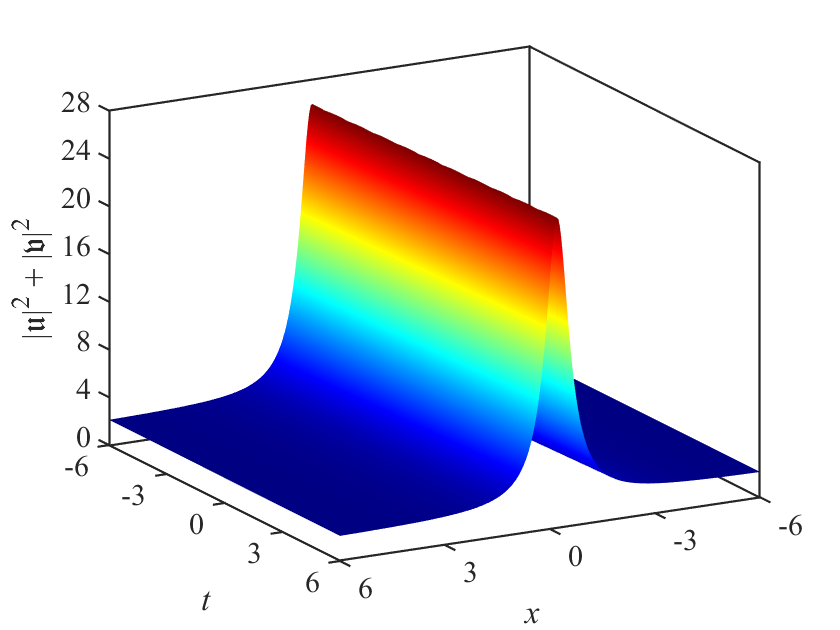}
		\vspace{0.1cm} 
		(a) 
	\end{minipage}
	\hfill
	\begin{minipage}[t]{0.45\columnwidth}
		\centering
		\includegraphics[width=\textwidth,height=6cm]{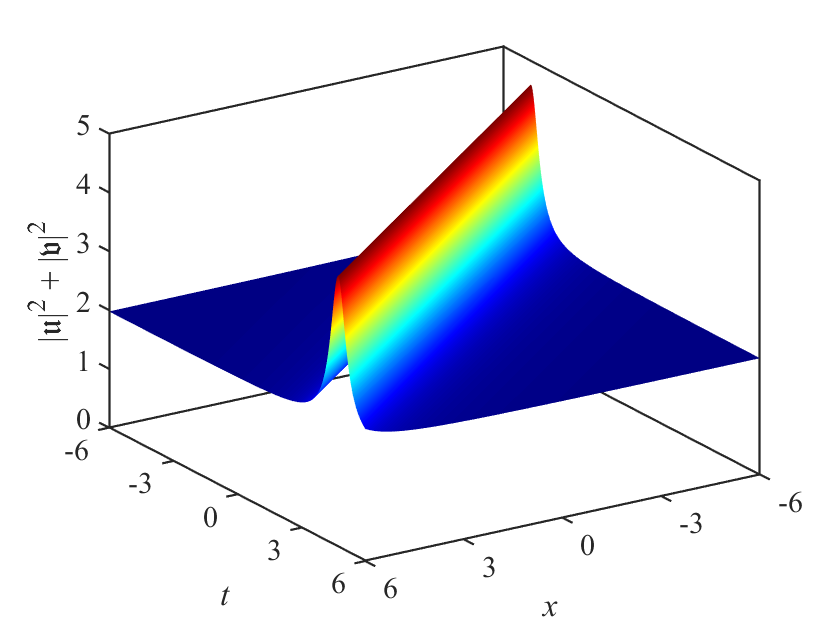}
		\vspace{0.1cm}
		(b) 
	\end{minipage}
	\caption{The same as in Figure \ref{FIG:9} but for the algebraic dark soliton with $c = c_+$ for (a) $\rho_{1}=0.5$, $\rho_{2}=-1$ and (b) $\rho_1 = 1.5$, $\rho_2 = 1$.}
	\label{FIG:11}
\end{figure}

Figures \ref{FIG:9} and \ref{FIG:11} show the spatial profiles 
of the algebraic dark soliton with $c = c_-$ and $c = c_+$. 
For the case $\rho_{1}=0.5$, $\rho_{2}=-1$ on panels (a), we have 
$c_- \approx -0.87$ and $c_+ \approx  -0.03$ and the exponentially decaying solutions may only exist for $c \in (c_-,c_+)$. 
For the case $\rho_1 = 1.5$, $\rho_2 = 1$ on panels (b), 
we have $c_- \approx 1.27$ and $c_+ \approx -4.04$ 
and the exponentially decaying solutions may only exist for $c \in (-\infty,c_+) \cup (c_-, \infty)$ since this case corresponds to 
$4 \rho_1^3 \rho_2^3 > (\rho_1^2 + \rho_2^2)^2$ in Proposition \ref{prop-velocity}. 


\subsection{Rogue wave}

The general expression for the single rogue wave was obtained in 
\cite{Baofeng1} for $N = 1$. For reader's convenience, we recover the 
explicit expression for the single rogue wave in the following proposition.

\begin{prop}
	\label{prop-rogue}
If $0<\rho_{1} \rho_{2} <1$, there exists only one solution of the bilinear equations \eqref{Hirota-bilinear} in the form:
\begin{align}
	\label{rogue wave}
	\begin{cases}
	f = f^+f^-+1,\\
	g = (\bar{f}^+- C_2 + {\rm i}D_2) (\bar{f}^- + C_2 + {\rm i}D_2 )+1, \\
	h = (\bar{f}^+-2) (\bar{f}^- +2) + 1,
	\end{cases}
\end{align}
where 
\begin{align}
\label{f-functions}
\begin{cases}
f^+ = (A_{1}+{\rm i}B)x+(A_{2}-{\rm i}B)t+C_{1}-{\rm i}D_{1}, \\
f^- = (A_{1}-{\rm i}B)x+(A_{2}+{\rm i}B)t-C_{1}-{\rm i}D_{1},
\end{cases}
\end{align}
with uniquely defined real coefficients $A_1,A_2,B,C_1,D_1,C_2,D_2$. 
The corresponding solution of the MTM system (\ref{MTM}) is uniquely defined in the form (\ref{rogue}) by (\ref{transfor}), (\ref{Hirota}), (\ref{rogue wave}), and (\ref{f-functions}) modulo the translational and rotational symmetry transformations. 
\end{prop}

\begin{proof} 
We verify that $f$, $g$, $h$ given by (\ref{rogue wave}) and (\ref{f-functions}) satisfy the four bilinear equations \eqref{Hirota-bilinear}. Substituting \eqref{rogue wave} and (\ref{f-functions}) into the first equation of the system \eqref{Hirota-bilinear} and gathering coefficients of the powers of $x$ and $t$, we obtain a system of equations for real coefficients $A_1$, $A_2$, $B$, $C_1$, $C_2$, $D_1$, and $D_2$. It follows from the explicit expressions
\begin{align*}
g_t + g_x &= 2 (A_1+A_2) (A_1 x + A_2 t + {\rm i}(D_1 + D_2)), \\
f_t + f_x &= 2 (A_1+A_2) (A_1 x + A_2 t - {\rm i} D_1), \\
h \bar{f} - g f &= (1 + \bar{f}^+ \bar{f}^-) (-4 - (2-C_2)(\bar{f}^+ - \bar{f}^-) 
- {\rm i} D_2 (\bar{f}^+ + \bar{f}^-) + C_2^2 + D_2^2)+\\
&\quad 4 {\rm i} g (D_1 (A_1 x+A_2 t)+B C_1 (x-t)),
\end{align*}
that ${\rm i} f(g_t + g_x) - {\rm i} g (f_t + f_x)$ is quadratic in $x$ and $t$, whereas $h \bar{f} - g f$ is cubic in $x$ and $t$. For the coefficients of the highest powers $x^3$ and $t^3$, the coefficients are 
	\begin{align*}
x^3 : \quad &		\frac{2{\rm i} \rho_{2} }{\rho_{1}} \left(B^{2}+A_{1}^{2}\right) \left[A_{1} (2 D_{1}-D_{2}) +B (2 C_{1} +C_{2}-2 )\right]=0, \\
t^3 : \quad &		\frac{2 {\rm i} \rho_{2} }{\rho_{1}} \left(B^{2}+A_{2}^{2}\right) \left[A_{2} (2 D_{1}-D_{2}) -B (2 C_{1} +C_{2}-2 )\right]=0.
	\end{align*}
Dividing by $\frac{2 {\rm i} \rho_{2} }{\rho_{1}}$ and adding together, we obtain 
	\begin{align*}
		(2D_{1}-D_{2})(A_{1}+A_{2})=0.
	\end{align*}
We assume $A_{1}+A_{2}\neq 0$ and $B \neq 0$ for nondegeneracy, after which the previous system of equations yields the first constraints:
	\begin{align}
	\label{relationD2}
		D_{2} = 2 D_{1}, \quad 		C_{2} = 2 (1- C_{1}).
	\end{align}
Substituting \eqref{relationD2} into the constant term of the first equation of the system \eqref{Hirota-bilinear}, we can obtain another constraint:
\begin{align}
	\label{relationD1}
D^2_1 = C_1 (1 - C_1).
\end{align}
After substituting the constraints \eqref{relationD2} and \eqref{relationD1} into the first equation of the system \eqref{Hirota-bilinear}, the coefficients of $x^2 t$ and $xt^2$ completely disappear. The coefficients of $x^2$, $t^2$ and $xt$ can be rewritten with the account of \eqref{relationD1} in the form:
	\begin{align*}
	x^2 : \quad &	\frac{\rho_2}{\rho_1} \left[ -2D^2_1 (A_{1}^{2}-B^{2}) - 2 (2C_1 - 1) D_{1} A_{1} B \right] \\
	& \qquad +  ( A_{1}+A_{2})[D_{1}  (A_{1}^{2}-B^{2})
	+ (2C_1 - 1)  A_{1} B]  = 0, \\
	t^2 : \quad &	\frac{\rho_2}{\rho_{1}}
 \left[-2D^2_1 (A_2^2 - B^2) + 2 (2C_1 - 1) D_1 A_2 B \right]  \\
 & \qquad + (A_{1}+A_{2}) [D_{1} (A_2^2 - B^2) - (2C_1 - 1)  A_2 B ]  = 0,\\
	xt: \quad & \frac{\rho_2}{\rho_1}  \left[ -4D^2_1(A_1 A_2 + B^2) + 2(2C_1 - 1)(A_1 - A_2)D_1 B \right] \\
	& \qquad +(A_1 + A_2)[ 2D_1(A_1 A_2 + B^2) - (2C_1 - 1)(A_1 - A_2)B] = 0,
\end{align*}
where we have divided each equation by the numerical coefficient $8$.
Adding the three equations together yields 
\begin{align*}
	(A_1+A_2)^2 \left[ D_1  (A_1+A_2) - \frac{2 \rho_2}{\rho_1} D_1^2 \right] =0.
\end{align*}
If $D_1  =  0$, then the system of equations for coefficients of $x^2$, $t^2$ and $xt$ imply for $A_1 + A_2 \neq 0$ and $B \neq 0$ that $C_1 = \frac{1}{2}$, which contradicts (\ref{relationD1}). Hence, $D_1 \neq 0$ and we have 
\begin{equation}
	\label{relationA1A21}
	A_1+A_2=\frac{2 \rho_2 D_1}{\rho_1},
\end{equation}
after which the system of equations for coefficients of $x^2$, $t^2$ and $xt$ 
is satisfied identically. After the constraints \eqref{relationD2}, 
\eqref{relationD1}, and \eqref{relationA1A21} are used, the coefficients of $x$ and $t$ in the first equation of the system \eqref{Hirota-bilinear} are satisfied identically. Hence, the first equation of the system (\ref{Hirota-bilinear}) is satisfied.

Next, we turn to the third equation of the system \eqref{Hirota-bilinear}. 
By using the constraints \eqref{relationD2} and \eqref{relationD1},  we obtain the coefficients of $x^2$, $t^2$, and $xt$ in the form:
\begin{align*}
	x^2 : \quad &2\rho_2^2 \left[ (C_1 - 1)(A_1^2 - B^2) - 2A_1 B D_1 \right]+(A_1 + A_2)[ D_1(A_1^2 - B^2) + 2C_1 A_1 B]   = 0, \\
	t^2 : \quad & 2\rho_2^2 \left[ (C_1 - 1)(A_2^2 - B^2) + 2A_2 B D_1 \right]+(A_1 + A_2)[ D_1(A_2^2 - B^2) - 2C_1 A_2 B]  = 0, \\
	xt : \quad & 4\rho_2^2 \left[ (C_1 - 1)(A_1 A_2 + B^2) + (A_1 - A_2) B D_1 \right]\\
	&\qquad + (A_1 + A_2)[2D_1(A_1 A_2 + B^2)  - 2C_1 B(A_1 - A_2)]  = 0.
\end{align*}
where we have divided each equation by the numerical coefficient $4$.
Adding the three equations together yields 
\begin{align*}
(A_{1}+A_{2})^{2} \left[ 2 \rho_2^2 (C_1 - 1) + (A_1 + A_2) D_1 \right]=0.
\end{align*}
By using \eqref{relationD1} and \eqref{relationA1A21}, this equation can be rewritten in the equivalent form:
\begin{align*}
\frac{2\rho_2}{\rho_1} (A_{1}+A_{2})^{2} (1-C_1) (C_1 - \rho_1 \rho_2) = 0.
\end{align*}
Since $A_1+A_2\neq 0$ and $C_1 \neq 1$ (due to $D_1 \neq 0$), we obtain the constraint:
\begin{align}
	\label{relationC1}
	C_1=\rho_1\rho_2,
\end{align}
after which the system of equations for coefficients of $x^2$, $t^2$ and $xt$ 
is satisfied identically. After the constraints \eqref{relationD2}, 
\eqref{relationD1}, \eqref{relationA1A21}, and \eqref{relationC1}  are used, the third equation of the system \eqref{Hirota-bilinear} is satisfied identically.

Next, we turn to the fourth equation of the system \eqref{Hirota-bilinear}. 
By using the constraints \eqref{relationD2} and \eqref{relationD1}, we obtain the coefficients of $x^2$, $t^2$, and $xt$ in the form:
\begin{align*}
	x^2: \quad & 2\rho_1^2 [ (A_1^2 - B^2)(4C_1 - 1)(C_1 - 1)  - 2A_1 B D_1(4C_1 - 3)] \\
	& \qquad + 2BC_1(A_1 A_2 - B^2) - D_1 \left[ A_1^2(A_1 - A_2) + B^2(3A_1 + A_2) \right] = 0, \\
	t^2: \quad & 2\rho_1^2 [ (A_2^2 - B^2)(4C_1 - 1)(C_1 - 1)  + 2A_2 B D_1(4C_1 - 3) ] \\
	& \qquad + 2BC_1(A_1 A_2 - B^2)  - D_1 \left[ A_2^2(A_1 - A_2) - B^2(A_1 + 3A_2) \right] = 0, \\
	xt: \quad & 4\rho_1^2 [ (A_1 A_2 + B^2)(4C_1 - 1)(C_1 - 1) + B D_1(A_1 - A_2)(4C_1 - 3) ] \\
	& \qquad + 2BC_1(A_1^2 + A_2^2 + 2B^2)  + 2D_1(A_1 - A_2)(B^2 - A_1 A_2) = 0.
	\end{align*}
where we have divided each equation by the numerical coefficient $4$. 
Adding the three equations together yields 
\begin{align*}
	\left(A_{1}+A_{2}\right)^{2} \left[ 2 \rho_{1}^{2} (4C_1 - 1)(C_1 - 1) + 2 B C_{1}-D_{1} \left(A_{1}-A_{2}\right)\right]=0.
\end{align*}
Since $A_1+A_2\neq 0$, we obtain the constraint:
\begin{align}
	\label{relation4}
2 \rho_{1}^{2}  (4C_1 - 1)(C_1 - 1) + 2 B C_{1}-D_{1} \left(A_{1}-A_{2}\right)=0.
\end{align}
Before solving the constraint (\ref{relation4}), we obtain the coefficient 
of the constant term in the fourth equation in the system \eqref{Hirota-bilinear}:
\begin{align*}
	4 \left(1-C_{1}\right) \left[2  \rho_{1}^{2}(C_{1}-1)+2 B C_{1}+D_{1} \left(A_{1}-A_{2}\right)\right]=0,
\end{align*}
Since $C_1 \neq 1$, this equation yields the constraint:
\begin{align}
	\label{relation5}
	2  \rho_{1}^{2}(C_{1}-1)+2 B C_{1}+D_{1} \left(A_{1}-A_{2}\right)=0.
\end{align}
Since $C_1 \neq 0$ (due to $D_1 \neq 0$), we add \eqref{relation4} and \eqref{relation5} together, divide by $4 C_1$, and obtain
\begin{align}
	\label{relation B}
	B = 2 \rho_{1}^{2} (1-C_{1}) = 2 \rho_{1}^{2} (1-\rho_1 \rho_2),
\end{align}
where we have used (\ref{relationC1}). Since $D_1 \neq 0$, we subtract \eqref{relation5} from \eqref{relation4}, divide by $2D_1$, and obtain
\begin{align}
	\label{relation A1A22}
	A_1-A_2=\frac{2 \rho_{1}^{2} (1-2C_1) (1-C_1)}{D_{1}}.
\end{align}
Combining \eqref{relationA1A21} and \eqref{relation A1A22}, and using 
\eqref{relationD1} and \eqref{relationC1}, we obtain the explicit expressions for $A_1$ and $A_2$:
\begin{align}
\label{relationsA}
	A_1 = \frac{(1-\rho_{1} \rho_{2}) (\rho_1^2 + \rho_2^2 - 2 \rho_1^3 \rho_2)}{D_1}, \quad 
	A_2=\frac{(1-\rho_{1} \rho_{2}) (\rho_2^2 - \rho_1^2 + 2 \rho_1^3 \rho_2)}{D_1}.
\end{align}
Furthermore, we have from (\ref{relationD2}) that $C_2 = 2(1-\rho_1 \rho_2)$ 
and $D_2 = 2 D_1$, where $D_1$ is given by 
\begin{equation}
\label{relationDplus}
D_1 = \pm \sqrt{\rho_1\rho_2(1 - \rho_1\rho_2)},
\end{equation}
where $\rho_1 \rho_2 (1- \rho_1 \rho_2) > 0$ due to $0 < \rho_1 \rho_2 < 1$. After the constraints \eqref{relation B} and \eqref{relationsA} are added to the previous constraints (\ref{relationD2}), (\ref{relationD1}), and (\ref{relationC1}), the fourth equation of the system \eqref{Hirota-bilinear} is satisfied identically.

All coefficients of the representation (\ref{rogue wave}) and (\ref{f-functions}) are defined. By using these definitions, we have confirmed that the second bilinear equation of the system \eqref{Hirota-bilinear} is satisfied identically.

It remains to prove that the two signs in (\ref{relationDplus}) return exactly the same solution for (\ref{rogue wave}) and (\ref{f-functions}). Transformation $D_1 \to -D_1$ yields $A_1\mapsto -A_1$, $A_2 \mapsto -A_2$, 
$B \mapsto B$, $C_1 \mapsto C_1$, $C_2 \mapsto C_2$, and $D_2 \mapsto -D_2$.  Substituting this transformation into \eqref{f-functions} yields
	\begin{align*}
		f^{+}&\mapsto (-A_{1}+{\rm i}B)x+(-A_{2}-{\rm i}B)t+C_{1}+{\rm i}D_{1}\\
		&=-[(A_1-{\rm i} B)x+(A_2+{\rm i} B)t-C_1-{\rm i} D_1]\\
		&=-f^{-},\\
		f^{-}&\mapsto (-A_{1}-{\rm i}B)x+(-A_{2}+{\rm i}B)t-C_{1}+{\rm i}D_{1}\\
		&=-[(A_1+{\rm i} B)x+(A_2-{\rm i} B)t+C_1-{\rm i} D_1]\\
		&=-f^{+}.
	\end{align*}
This transformation does not change the solution \eqref{rogue wave} 
since $f \mapsto f$, $g \mapsto g$, and $h \mapsto h$. Hence, the two signs in (\ref{relationDplus}) return only one solution for (\ref{rogue wave}) with (\ref{f-functions}).
\end{proof}

\begin{figure}[htb!]
	\centering
	\includegraphics[width=0.5\columnwidth,height=6cm]{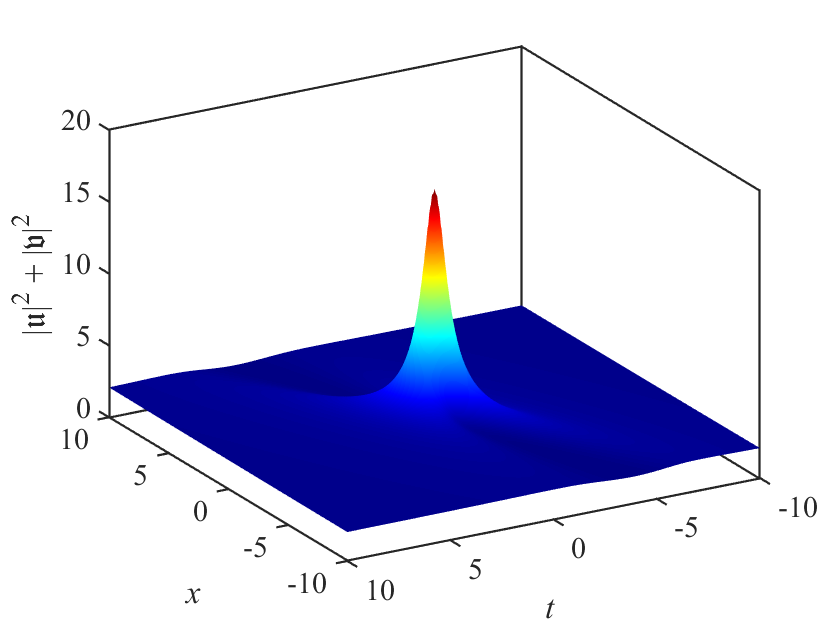}
	\includegraphics[width=0.45\columnwidth,height=6cm]{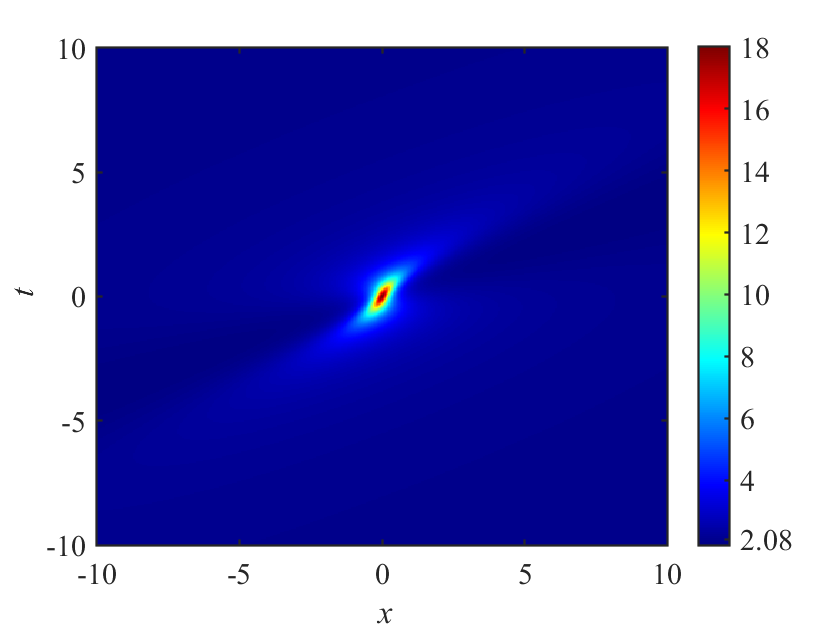}
	\caption{The total intensity $|\mathfrak{u}|^2+|\mathfrak{v}|^2$ versus $(x,t)$ of the rogue wave for $\rho_{1}=1.2$ and $\rho_{2}=0.8$: the solution surface (left) and the density plot (right).}
	\label{FIG:25}
\end{figure}

In order to address the decay of rogue waves and the maximal intensity (\ref{max-elevation-rogue}) in Theorem \ref{theorem-soliton}, we rewrite the expressions for $f$, $g$, and $h$ in the explicit form:
\begin{align}
\label{rogue wave1}
	\begin{cases}
	f&= (A_1 x+A_2 t)^2+B^2 (x-t)^2 - 2 {\rm i} [D_1 (A_1 x+A_2 t)+B C_1 (x-t)] +1-C_1,\\
	g&= (A_1 x+A_2 t)^2+B^2 (x-t)^2 + 2 {\rm i} [3 D_1 (A_1 x+ A_2 t) + B (3 C_1-2) (x-t) ] \\
	& \quad +1 - (3 C_1-2)^2 - 9 D_1^2, \\
	h&= (A_1 x+A_2 t)^2+B^2 (x-t)^2 + 2 {\rm i} [D_1 (A_1 x+A_2 t) + B (C_1-2) (x-t) ] \\
	& \quad +1 - (C_1-2)^2 -D_1^2,
	\end{cases}
\end{align}
where we have used relations (\ref{relationD2}). It follows from \eqref{rogue wave1} that
\begin{align*}
	g-\bar{f} &= -4 (1-C_1)+4 {\rm i} [D_1 (A_1 x+A_2 t)- B (1-C_1) (x-t)],\\
	h-f &= -4 (1-C_1)+4 {\rm i} [D_1 (A_1 x+A_2 t)- B (1-C_1) (x-t)],
\end{align*}
where we have used relations (\ref{relationD1}). Since $A_1 + A_2 \neq 0$ and $B \neq 0$, we immediately obtain that 
\begin{equation*}
|\hat{u}-1| = \frac{|g-\bar{f}|}{|\bar{f}|} =\frac{|h-f|}{|f|} = |\hat{v}-1|
\end{equation*}
with the algebraic decay given by  
$$
|\hat u(x,t) - 1| = |\hat v(x,t)-1| = \mathcal{O}((|x|+|t|)^{-1}) \quad \mbox{\rm as} \;\; |x| + |t| \to \infty.
$$ 
Furthermore, since $|f|^2$, $|g|^2$, $|h|^2$ are quartic polynomials in $x$ and $t$, the extremal value of $|\hat{u}(x,t)|^2 + |\hat{v}(x,t)|^2$ occurs at $(x,t) = (0,0)$, for which we obtain from (\ref{rogue wave1}) that 
\begin{align*}
(x,t) = (0,0): \quad 
\begin{cases}
f&= 1-\rho_1\rho_2,\\
g&= -3(1-\rho_1 \rho_2), \\
h&= -3(1-\rho_1 \rho_2), 
\end{cases}
\end{align*}
where we have used relations (\ref{relationD1}) and (\ref{relationC1}). The maximum intensity is attained at $(x,t) = (0,0)$:
\begin{align*}
\max_{(x,t)\in\mathbb R^2}
\bigl(|\hat u(x,t)|^2+|\hat v(x,t)|^2\bigr) = 18.
\end{align*}
This expression recovers (\ref{max-elevation-rogue}). It is remarkable that the maximal normalized intensity is independent of parameters $\rho_1$, $\rho_2$, which are only needed to satisfy $0 < \rho_1 \rho_2 < 1$ for the existence of the rogue wave in Proposition \ref{prop-rogue}. Figure \ref{FIG:25} shows the solution surface and the density plot of a representative rogue wave solution for $\rho_{1}=1.2$ and $\rho_{2}=0.8$.

\section{Conclusion}
\label{sec-conclusion}

Here, we summarize our findings and emphasize the open questions beyond this study.

We developed a comprehensive analysis of the spectral stability of the nonzero wave background in the MTM system and identified the parameter regions, in which the background is stable or unstable (Theorem \ref{theorem-stab}). In the case of instability, we found the parameter regions, where the instability is modulational, or side-band, or full-band. Furthermore, outside the modulational instability region, we constructed exactly two algebraically decaying dark 
solitons that propagate with two distinct wave speeds, whereas inside the modulational instability region, we constructed only one distinct fundamental rogue wave (Theorem \ref{theorem-soliton}). We connected the wave speeds of the two algebraically decaying dark solitons with the end points in the admissible range of the wave speed for the exponentially decaying dark solitons  and pointed out that if the background is spectrally unstable with respect to side-band or full-band instabilities, then the dark solitons are also spectrally unstable in the time evolution problem. 

It is not clear if the rational solutions for the algebraic dark solitons can be extended beyond the one-soliton solutions. It seems to be possible to construct a two-soliton rational solution with each algebraic soliton travelling with one of the two distinct wave speeds. However, it is an open problem if two-soliton and multi-soliton rational solutions can be constructed to describe slow interactions of identical dark solitons traveling with the same wave speed. This question is related to the open question whether multiple eigenvalues of the spectral problem for the MTM system at the nonzero background may exist in the same way as they exist for the spectral problem at the zero background \cite{Han-2024,LiPelin2025,ZhaoPelin2026}. We note that the hierarchy of multiple rogue wave solutions has already been constructed for the MTM system at the nonzero background in the modulational unstable regime \cite{Baofeng1,He}.

\vspace{0.25cm}

{\bf Acknowledgements.}  The work of Z. Zhao was conducted during their PhD studies while visiting McMaster University with the financial support from the China Scholarship Council. The work of D.E. Pelinovsky was supported by the National Natural Science Foundation of China (No. 12371248).

\bibliographystyle{plain}

\end{document}